\documentclass[12pt]{iopart}
\usepackage{amssymb}
\usepackage{graphicx}

\renewcommand*\contentsname{Quantum correlations in ``classical'' light}
\begin{document}

\title[Quantum correlations in ``classical'' light]{Quantum correlations in separable multi-mode states and in classically entangled light}

\author{N. Korolkova$^1$ and G. Leuchs$^2$}

\address{$^1$ School of Physics and Astronomy, University of St. Andrews, North Haugh, St. Andrews, Fife, KY16 9SS, Scotland}
\address{$^2$ Max Planck Institute for the Science of Light, Staudtstra{\ss}e 2, 91058 Erlangen, Germany;
Institute of Optics, Information and Photonics,
University of Erlangen-Nuremberg, Staudtstra{\ss}e 7/B2, Erlangen, Germany}
\ead{$^1$nvk@st-andrews.ac.uk; $^2$gerd.leuchs@mpl.mpg.de}
\vspace{10pt}

\begin{abstract}
In this review we discuss intriguing properties of apparently classical optical fields, that go beyond purely classical context and allow us to speak about quantum characteristics of such fields and about their applications in quantum technologies. We briefly define the genuinely quantum concepts of entanglement and steering. We then move to the boarder line  between classical and quantum world introducing quantum discord, a more general concept of quantum coherence, and finally a controversial notion of classical entanglement. To unveil the quantum aspects of often classically perceived systems, we focus more in detail on quantum discordant correlations between the light modes and on nonseparability properties of optical vector fields leading to entanglement between different degrees of freedom of a single beam. To illustrate the aptitude of different types of correlated systems to act as quantum or quantum-like resource,  entanglement activation from discord, high-precision measurements with classical entanglement  and quantum information tasks using intra-system correlations are discussed. The common themes behind the versatile quantum properties of seemingly classical light are coherence, polarization and inter and intra--mode quantum correlations.
\end{abstract}
\maketitle
\tableofcontents

\section{Introduction}

Quantum information science carries the potential for radically new means of computation, communication and information processing through the use of novel devices exploiting the laws of quantum mechanics. Coherence is a key notion here, being the underlying fundamental resource for all quantum information technologies. For instance, computation speed-up in the quantum realm comes from parallelism that stems from quantum coherent superposition states. Furthermore, all entanglement-enabled technologies in information processing and high precision measurements are emerging from coherence in its most general meaning, that is, from the correlations between physical quantities of quantum systems, from single quanta to optical modes, and their ability to interfere in a deterministic manner. However, it is not yet well understood how quantum correlations, or ultimately coherence, impacts on the computational performance and complexity of tasks that can be implemented and how to efficiently protect such systems from the environment. Studying decoherence and noise effects in combination with coherence and correlations provides vital insights for radically new technologies. The notion of quantum correlations is also wider in a more general, more realistic scenario of noise-affected quantum states, quantum statistical mixtures of quantum states. In mixed states, multi-partite quantum correlations are not equivalent to entanglement and also separable, seemingly classical systems manifest quantum features. The purpose of this review is to introduce the reader to quantum resources hidden in systems, which are not necessarily obviously quantum but exhibit different coherent properties with potential applications in quantum technologies. We start with entanglement and steering, as ultimate quantum resources, then move on to quantum correlations in separable mixed states, as described by quantum discord; and finally introduce ``classical'', or more precisely, quantum intra-system entanglement in vector fields in optics.

\section{Definitions}
\subsection{Quantum correlations in nonseparable and separable states}
\subsubsection{Quantum entanglement, quantum steering and beyond}
Consider a composite system $AB$ described by the density matrix operator $\hat \rho_{AB}$.
When speaking about such systems, we think in the first place of entangled, nonseparable states as quantum. What is {\it quantum entanglement}?
A state is said to be {\it entangled} if it is {\it nonseparable}, that is, if and only if it cannot be fully written as a convex decomposition of the product states of the subsystems $A$ and $B$ \cite{werner}:
\begin{eqnarray} 
\hat \rho_{AB} \neq \sum_{i} p_{i} \hat \rho_{Ai} \otimes \hat \rho_{Bi},
\label{ent}
\end{eqnarray}
where $Ai$, $Bi$ denote different possible decompositions of the global state $AB$.
Entangled state is a quantum coherent superposition state, as is lucidly reflected in the form of the four maximally entangled pure states, the Bell states, forming a complete basis for description of a general entangled state. For a two-level system with possible states $\vert \uparrow \rangle, \vert \downarrow\rangle$ the Bell states read:
\begin{eqnarray}
\vert \Psi^{\pm} \rangle _{AB} = \frac{ 1}{\sqrt 2} \Big ( \vert \uparrow \rangle _A\vert \downarrow\rangle _B \pm \vert \downarrow\rangle _A \vert \uparrow\rangle _B \Big ), \\
\vert  {\Phi^{\pm}} \rangle _{AB} = \frac{ 1}{\sqrt 2}\Big (\vert \uparrow\rangle _A\vert \uparrow\rangle _B \pm \vert \downarrow\rangle _A \vert \downarrow\rangle _B\Big ). \label{Bell}
\end{eqnarray}
Entanglement is a manifestation of very strong and non-local correlations stemming from the fact that the two entangled subsystems are described by a common global wavefunction. As clearly seen in the example above, it is a coherent superposition of two possible states of the global system, $\vert \uparrow \rangle _A\vert \downarrow\rangle _B$ and $\vert \uparrow \rangle _B\vert \downarrow\rangle _A$. Measurement on one subsystem of the entangled state instantly affects the state of the other subsystem, leading to highly correlated measurement outcomes on both (eventually remote) subsystems $A$ and $B$. 

The notion of {\it quantum steering} \cite{Wiseman_2007,Jones_2007,He_2015} has emerged in the recent years to highlight one of the properties inherent to entangled states: if a measurement is performed on a subsystem $A$, different outcomes can lead to different states of another subsystem $B$. That is, measurement on $A$ steers the state of $B$.
Quantum steering is not identical with entanglement. For pure states, all quantum states that exhibit steering are entangled but not all entangled states are steerable. The idea of steering goes back to the work of  Schr\"odinger \cite{schroedinger} on discussion of the Einstein Podolsky Rosen (EPR) paradox \cite{EPR}. There he introduced the term ``entanglement'' to describe the states with EPR-like nonlocal correlations (like defined in our Eq.~(\ref{ent})) and used ``steering'' to describe the effect of the measurement of one part of the EPR state  on the quantum state of the other part. Formally steering has been defined only in 2007 \cite{Wiseman_2007}. In the hierarchy of entangled states \cite{Wiseman_2007}, Bell nonlocal states are at the highest level. A state is called Bell-nonlocal iff there exist a measurement  set that allows Bell nonlocality to be demonstrated. Stricktly weaker is nonseparability or entanglement as defined in Eq.~(\ref{ent}). Both concepts, Bell-nonlocality and entanglement, are symmetric between $A$ (Alice) and $B$ (Bob). Steering is inherently asymmetric. Steering is connected to whether Alice, by her choice of measurement on $A$, can collapse Bob's system $B$ into different type of states in different ensembles. Concept of steering place decisive role in quantum communication protocols like teleportation, secrete sharing, quantum key distribution and in general in secure quantum networks.

However, of course, also separable states can be distinctly quantum. The simplest manifestation of this is the
non-compatibility of observables, which is as true for separable quantum systems, as for more sophisticated and exotic states.
The quantum measurements will generally alter these states. Non-orthogonal states cannot be discriminated deterministically and exactly. Hence, the correlations present in separable states can very well have quantum nature. Along the same lines, as local measurements can affect the general quantum separable states, they therefore can exhibit some kind of steering \cite{xiang-adesso,Pusey-PhD}. The structure of steering can be more subtle in separable states and is known as ``incomplete'' or ``surface steering'' \cite{Pusey-PhD}.

{\it Quantum discord} is one of the most used quantifies of quantum correlations beyond entanglement
\cite{Zurek,Vedral_01} and is inherently asymmetric, like quantum steering. This and other measures of non-classical correlations where specifically designed to capture more general signatures of quantumness in composite systems. As already mentioned, they are connected largely to the fact that local measurements generally induce some disturbance to quantum states, apart from very special cases in which those states admit a fully classical description. This fact allows us to answer the first question, which naturally arises when speaking about general bi-partite quantum states.  How to distinguish between classical and quantum correlations? In its simplest version, the answer is: a classically correlated state can be measured and determined without altering it. For pedagogical introductions on this topic see \cite{modi14,Vedral_17}. For detailed technical reviews on quantum discord, other measures and applications of quantum correlations, and on the classical-quantum boundary for correlations see Modi {\it et al} \cite{modi-review}, Adesso {\it et al} \cite{adesso-review} and Streltsov \cite{streltsov-review}.

\subsubsection{Definitions of quantum  discord}

As we have seen in the previous section, the notion of quantum measurement and measurement-induced disturbance plays a crucial role in the concept of quantum correlations. This is reflected in all the definitions of quantum discord and other measures of non-classical correlations. However emphasis can be put on different aspects of measurement process or of nonclassicality notion. One can restrict oneself to a particular set of measurements. Alternatively, one can search for an optimized subset of measurements, looking for gaining as much information as possible (optimizations based on supremum), or for inducing the least disturbance (infimum optimizations). As an alternative approach, geometrical or distance measures can also be introduced \cite{modi-review}, which quantify the quantumness of state in question by assessing its distance to the relevant classical reference state.

Quantum discord has first been introduced in terms of von Neumann entropies, which included optimization over certain class of measurements. Let us start though with a much simpler definition, which clearly highlights the essence of notion ``quantum correlations''. \\ \\
{\it Definition I.}
A concise and tangible definition of quantum discord has been given by Modi \cite{modi14}: \\
A state is said to be discordant if and only if it cannot be fully determined without disturbing it with the aid of local measurements and classical communication:
\begin{eqnarray}
\hat \rho_{AB} \neq \sum_{ab} \vert ab \rangle \langle ab \vert \hat \rho_{AB} \vert ab \rangle \langle ab \vert 
= \sum_{ab} p_{ab} \vert a \rangle \langle a \vert \otimes \vert b \rangle \langle b \vert
\label{discord-modi}
\end{eqnarray}
where $\{ \vert ab \rangle\}$ forms an orthonormal basis, $\langle ab \vert a^\prime b^\prime \rangle = \delta_{aa^\prime}\delta_{bb^\prime}$.  This definition is formulated in the same fashion as the definition of nonseparable states
(\ref{ent}) and is quite insightful, also in the context of ${P}$-classicality and ${C}$-classicality discussed later in this review.
The key feature of quantum correlations is a sensitivity of quantum correlated state to measurements on a subsystem.
\\ \\
{\it Definition II.}
Historically, the first definition formulated for quantum discord has been given in the language of the von Neumann entropies (see original papers \cite{Zurek,Vedral_01} and  brief introductory review \cite{Vedral_17}).
For two systems $A$ and
$B$ described by the density matrix operator $\hat \rho_{A}$, $\hat \rho_{B}$, quantum discord is defined as the difference 
\cite{Zurek},
\begin{eqnarray}\label{discord}
\mathcal{D}^\leftarrow({AB}) &=& {\cal I}({AB})
-\mathcal{J}^\leftarrow({AB}),
\end{eqnarray}
between quantum mutual information ${\cal I}({AB})={\cal
S}(\hat \rho_{A})+{\cal S}({\hat \rho_B})-{\cal S}(\hat \rho_{AB})$ encompassing all
correlations present in the system, and the one-way classical
correlation 
\begin{eqnarray}
\mathcal{J}^\leftarrow({AB})={\cal S}(\hat \rho_{A})-\inf_{\{\hat\Pi_i\}} {\cal H}_{\{\hat\Pi_i\}}(A|B), 
\end{eqnarray}
which is operationally related to the amount of perfect classical
correlations which can be extracted from the system
\cite{Devetak_04}. Here, ${\cal S}(\hat \rho)=-\mbox{Tr}(\hat \rho\ln\hat \rho)$ is the von Neumann entropy of
the respective state, ${\cal H}_{\{\Pi_i\}}(A|B){\equiv}\sum_{i}p_i{\cal
S}(\hat \rho^i_{A|B})$ is the
conditional entropy with measurement on $B$, and the infimum is
taken over all possible measurements $\{\hat\Pi_i\}$. As you can easily see, this definition points in the same direction: it assesses the amount of information about one subsystem gained upon measurement of the other subsystem versus the disturbance introduced by the measurement (see also Modi {\it et al} \cite{modi-review}). In other words, quantum discord tell us how much we can reduce the entropy of one subsystem by measuring the other.

\subsection{Quantum coherence} \label{qu-coh} 
Quantum correlations beyond entanglement as captured by quantum discord can be
understood as a certain type of classical correlations upheld with quantum coherence (including quantum coherent superpositions) at the level of individual subsystems \cite{Vedral_17}. Hence the research on discord has led to the research on quantifying {\it quantum coherence}, embracing different areas of physics where quantum coherence can represent a resource (for a lucid colloquim-style review see \cite{coherence}).

Coherence is basis dependent and the choice of the reference basis in normally determined by the physical problem in question. The definition of quantum coherence works in the same way as that of entanglement: we have defined separable quantum states and all other states that are nonseparable are entangled. The density matrices diagonal in the particular, chosen reference basis are called {\it incoherent} if:
\begin{eqnarray}
\hat \rho 
= \sum_{a=0}^{d-1} p_{a} \vert a \rangle \langle a \vert ,
\label{incoh}
\end{eqnarray}
where $d$ is the dimension of the corresponding Hilbert space \cite{coherence}. General multipartite incoherent states are defined as convex combinations of such incoherent pure product states \cite{streltsov-review,qu-coh}. States which are not of the form (\ref{incoh}) exhibit some form of {\it quantum coherence} \cite{qu-coh} and it is not possible to gain full information about such states without some form of penalty - the feature which we already highlighted when speaking about quantum discord. 
As stated in \cite{coherence}, only incoherent states are accessible free of charge.

\subsection{Classical entanglement}

{\it Classical entanglement} refers to quantum correlations between different degrees of freedom of one and the same physical system \cite{spreeuw,eberly2011}. In optics, the nonseparable mode function of some specifically shaped vector fields (vector beams) is mathematically equivalent to that of maximally entangled Bell states of two qubits (\ref{Bell}).
In this way, polarization theory can be reformulated as entanglement analysis and the tools from entanglement theory can be efficiently applied in polarization metrology. This new vision of polarization theory is best reflected in \cite{eberly2011}: ``polarization is a characterization of the correlation between the vector nature and the statistical nature of the light field''. We will discuss this concept and its applications in metrology in the last section of this review. 

\subsection{Quantumness in separable states and nonseparability of different properties of single system}
In what follows, we will not deal with such established (and obviously quantum) resources as quantum entanglement and quantum steering. Our main interest will be light fields, that appear merely classical at the first glance. The field of general quantum correlations is getting very broad (see, e.~g., \cite{lectures-gen-cor}) and we therefore concentrate in this review on two representative research areas. In sec.~\ref{discord-sec} we discuss in detail quantum correlations in mixed separable multi-mode optical fields and pay particular attention to the general correlations as measured by quantum discord. This is a vivid example of quantumness present in inter-system correlations in a composite but fully separable global system. Sec.~\ref{pol-cl-ent} unveil quantumness of intra-system correlations of otherwise classical beams. Such quantum entanglement present between different degrees of freedom (and thus local) has first been perceived much as a mathematical curiosity rather than a quantum resource, hence the name ``classical entanglement''. In the case of quantum discord, separability of two spatially separated systems has been concealing the quantum character of correlations between them. For classical entanglement, principle spatial inseparability of the entangled degrees of freedom impeded to embrace quantumness of such fully nonseparable system. We therefore find these two examples particularly inspiring in discovering new hidden quantum potential in very familiar and on the surface completely classical systems, and present here their more detailed treatment, starting from discordant correlations.

\section{Quantum correlations in mixed separable states of light.  }\label{discord-sec}
Discord has been defined and studied first for two-level systems and there is an extensive literature covering this subject. It is, however, very illuminating to turn to bosonic systems.  Bosonic systems have been a major framework to study the quantum-classical boundary over the last century. Even more instructive is to use optical modes as bosonic
objects to study correlations due to impressive breakthroughs achieved in photonics, quantum optics and photonic-based quantum information, combined with relative simplicity of the corresponding systems.

Quantum states of readily available photonic systems are in most cases described by the so-called Gaussian states.
Some considerations presented below are valid also for general bosonic systems, including such highly non-Gaussian states as single photons, photon number states, etc. Nevertheless,  many conceptually important features with respect to correlations, their measures, and their quantum or classical nature are distinctly tractable for Gaussian systems, which allows for relatively simple experimental demonstrations and elegant mathematical description (see, e.~g., \cite{reviewGauss}).

\subsection{Gaussian quantum discord}
\begin{center}
\begin{figure}[h]
\begin{center}
\includegraphics[width=9cm]{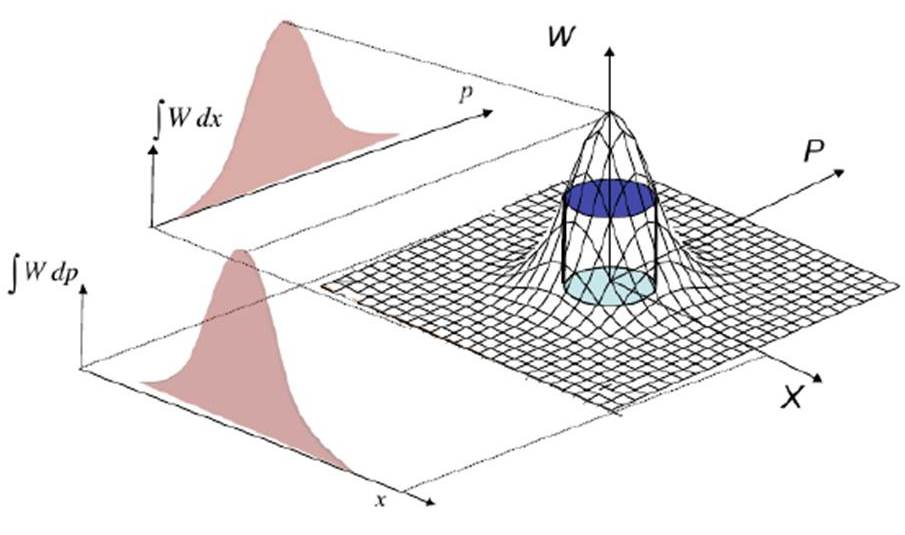} \end{center}
\caption{ (color online). Gaussian states are the states described by the Gaussian-shaped Wigner function, a quasi phase-space probability distribution, which marginal distributions are true probability distributions for position and momentum (or position and momentum quadratures for optical modes).}
\label{Gauss}
\end{figure}
\end{center}

Gaussian states are quantum states of systems in
infinitely-dimensional Hilbert space, e.~g., light modes, which
possess a Gaussian-shaped Wigner function \cite{wigner} (Fig.~\ref{Gauss}) and hence are completely described by the first and second moments of the respective probability distribution.  Each optical mode can be described by the apmlitude $\hat x= \frac{\hat a^\dagger + \hat a}{2}$ and phase $\hat p= \frac{\hat a^\dagger - \hat a}{ 2i}$ quadrature operators, which physically correspond to the real and imaginary parts of the electric field. Here $\hat a, \hat a^\dagger$ are the bosonic annihilation and creation operators, $\left [ \hat a, \hat a^\dagger \right ] = 1$. Up to the $\hbar$ factor, they have the same operator algebra as position and momentum operators, and consequently also same commutation relation, and therefore are sometimes called position $\hat x^{\theta = 0}=\hat x$ and momentum $\hat x^{\theta = \pi/2} = \hat p$ quadratures, where quantities labeled with $\theta$ correspond to the so called generalized quadratures 
$\hat x^{\theta}=({\hat a^\dagger \exp{(-i\theta)}+ \hat a\exp{(i\theta)}})/{2}$ with all possible pairs of conjugate variables in phase space scanned through an angle $\theta$ (see, e.g., ~\cite{wigner}).
Correlations carried by a Gaussian state of two modes $A$ and
$B$ are then completely characterized by $\gamma$, the covariance matrix
(CM)  \cite{reviewGauss}. Its entries are all second order moments $\langle \hat x^{\theta_i}_A x^{\theta_j}_B\rangle$,  $i, j=1,2$ where  $\theta_1 =0$, $\theta_2 =\pi/2$.

{\it Gaussian quantum discord} \cite{paris,adesso} is defined by Eq.~(\ref{discord}), where the
minimization in $\mathcal{J}^\leftarrow({AB})$ is restricted to
Gaussian measurements, $\{\hat\Pi_i^G\}$:
\begin{eqnarray}
\mathcal{D}^\leftarrow({AB}) &=& {\cal I}({AB})
-\mathcal{J}^\leftarrow({AB}) \nonumber \\  &=&{\cal S}(\hat \rho_{B}) - {\cal S}(\hat \rho_{AB})+\inf_{\{\hat\Pi_i^G\}} 
\sum_{i}p_i{\cal S}(\hat \rho^i_{A|B}).
\label{Gauss-discord}\end{eqnarray}
All non-product bi-partite Gaussian states
have been shown to have non-zero Gaussian discord \cite{adesso,Mista_14}. Gaussian discord carried by the state can be determined from $\gamma$ using the analytic formula derived in \cite{adesso,paris}. The Gaussian discord coincides with unrestricted discord (\ref{discord}) for states considered here \cite{Pirandola_14} which
confirms the relevance of its use.

\subsection{Quantifiers of quantum correlations and the optimal measurements}

Quantum discord is not the only suitable correlation quantifier; several other quantifiers has been introduced, all linked to a particular choice of the optimal measurement for a given quantum state under consideration. A good account of different measures is given in \cite{modi-review}. Here, we would mainly like to give a flavour of what is behind the different measures and what should guide you in choosing the one, most appropriate for your task. For comparison purposes, we have chosen the Measurement-induced disturbance (MID), $\mathcal M$, introduced by Luo \cite{Luo} and the Gaussian ameliorated  Measurement-induced disturbance (Gaussian AMID), $\mathcal {A^G}$, developed on its basis~\cite{gamid}. We compare them to the two-way Gaussian discord,
\begin{equation}
\mathcal{D}^\leftrightarrow(\hat{\rho}_{AB})=
\max\{\mathcal{D}^\leftarrow(\hat{\rho}_{AB}),
\mathcal{D}^\rightarrow(\hat{\rho}_{AB})\},
\nonumber
\end{equation}
for a large set of randomly generated Gaussian states.

\noindent
{\it Measurement-induced disturbance (MID).} MID has been introduced by Luo \cite{Luo}.  For a given Gaussian state the MID is a gap between its quantum mutual information — quantifying
the total correlations — and the classical mutual information of outcomes of local Fock-state detections:  
\begin{equation}
{\cal M}(\hat{\rho}_{AB})={\cal I}({AB})-
{\cal I}(A:B),\nonumber
\end{equation}
\begin{equation}\label{MI}
{\cal I}({AB})={\cal S}(\hat{\rho}_{A})+{\cal S}
(\hat{\rho}_{B})-{\cal S}(\hat{\rho}_{AB}), \qquad {\cal I}(A:B)=H(A)+H(B) - H(A,B),
\nonumber\end{equation}
where ${\cal I}(A:B)$ is the classical mutual information (of outcomes of local Fock measurements) and $H(x)$ is the Shanon entropy.
MID captures a specific type of non-Gaussian classical correlations in the state.

\noindent
{\it Gaussian ameliorated Measurement-induced disturbance (Gaussian AMID).}
The Gaussian AMID \cite{gamid} is a gap between the quantum mutual information and the maximal 
classical mutual information that can be obtained by local Gaussian measurements, 
the latter quantifying the maximum classical correlations that can be extracted from
the state by local Gaussian processing:
\begin{equation}
{\cal A}^{G}(\hat{\rho}_{AB})={\cal I}({AB})
-{\cal I}_{c}^{G}(\hat{\rho}_{AB}), \qquad
{\cal I}_{c}^{G}(\hat{\rho}_{AB})=\sup_{\{\hat{\Pi}_{A}^G\otimes
\hat{\Pi}_{B}^G\}}{\cal I}(A:B)
\nonumber
\end{equation}

\begin{center}
\begin{figure}[t]
\begin{center}
\includegraphics[width=15cm]{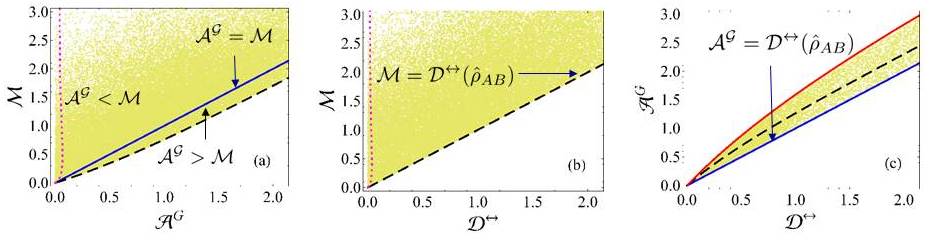} \end{center}
\caption{ (color online). Comparison between (a) MID versus Gaussian AMID, (b) MID versus two-way Gaussian discord, and
(c) Gaussian AMID versus two-way Gaussian discord, for $10^5$
randomly generated mixed two–mode Gaussian states (yellow dots building the background). Dashed black curve: pure
two–mode squeezed states, TMSV;  Red dotted line: thermal squeezed state; Blue solid line: boundary at which the compared measures coincide.  In panel (c) the blue boundary correspond to TMSV in the limit of infinite squeezing, the red solid boundary to thermal squeezed state. All the quantities plotted are dimensionless. Modified from \cite{gamid}. }
\label{compare}
\end{figure}
\end{center}
\noindent
{\it Gaussian discord, MID and Gaussian AMID compared for Gaussian states.}
The comparison of these three measures highlights nicely the importance of a choice of measurement set for assessing the quantum nature of a state (or for inferring information about one subsystem by 
performing measurements on another, correlated one).
Figure~\ref{compare} compares the three different correlation measures for the same set of states, randomly generated mixed two-mode Gaussian states \cite{gamid}. There are some special cases among them, such as pure entangled state two-mode squeezed vacuum (TMSV) (dashed black curve) \cite{wigner}; or thermal squeezed state (red dotted line) 
\cite{wigner}, important for a wide range of applications.

Fig.~\ref{compare} (a,b) clearly show, that MID is typically very loose and often overestimates the amount of quantum correlations. However, in Fig.~\ref{compare}~(a), in the segment between solid and dashed line ${\cal A}^{\cal G} >{\cal M} $. This means, that non-optimized non-Gaussian measurements (as in ${\cal M}$) reveal quantumness more accurately for this particular subset of mixed Gaussian states compared to optimized Gaussian POVM measurements used in Gaussian AMID. The region includes the pure TMSV states (cf. the dashed line corresponding to TMSV). We see that there is a certain threshold value beyond which the Gaussian POVMs are no longer optimal for the AMID, and non-Gaussian measurements such as photon counting (via MID) provide a more accurate result, culminating in the extreme case of pure states where those specific measurements are globally optimal.  It highlights the importance of non-Gaussian measurements in certain instances, revealing that  for correct quantification of (non)classical 
correlations in Gaussian states non-Gaussian processing might be in order.

When compared to quantum discord (Fig.~\ref{compare}~(b)), MID  almost always overestimate amount of quantum correlations. Photon counting measurements provide a very loose upper bound to quantum discord. Finally, we see that Gaussian AMID and quantum discord are intimately related (Fig.~\ref{compare}~(c)). Gaussian AMID admits upper and lower
bounds at a given value of the two-way Gaussian discord. The lower (blue online) boundary in panel (c) accommodates states for which the two quantifiers give identical prescriptions for measuring quantum correlations. These
are states with covariance matrix $\gamma$ in standard form, and correspond to the TMSV in the limit of infinite squeezing, i. e., pure maximally entangled state. Gaussian AMID is particularly accurate for mixed and strongly correlated states, such as thermal squeezed state, displayed here as an upper solid red line to the shaded yellow segment.

The considerations above illustrate very well the link between a suitable correlation quantifier and the optimal measurement for a particular quantum state under consideration. There is also a wide range of 
geometrical and distance measures, which are often much easier to compute that the other measures mentioned here. 
For a good account  of computability of these measures versus their reliability see \cite{adesso-geometric}.

\noindent{\it Von Neumann or R\'{e}nyi entropy for Gaussian states?}
Is the von Neumann entropy in the definition (\ref{Gauss-discord}) the most natural of entropies to be used in entropic quantum correlation measures, for example, for Gaussian states? It has been shown in \cite{sampling} that it is actually the R\'{e}nyi-2 entropy, that arises naturally from phase-space sampling for Gaussian states. 

The R\'{e}nyi entropy was introduced as a generalisation of the usual concept of entropy and in particular the classically implemented Shannon entropy \cite{orgpaper},
\begin{equation}
H_\alpha({P})=\frac{1}{1-\alpha}\log_2 \left(\sum_{k=1}^n p_k^\alpha\right),
\label{classicalrenyi}
\end{equation}
where ${P}$ is a probability distribution ${P}=[p_1,\ldots,p_k]$. Shannon's measure of entropy is the limiting case of $\alpha\rightarrow 1$ of Eq. (\ref{classicalrenyi}) obtained by L'H\^{o}pital's rule as
\begin{equation}
H_1({P})= \sum_{k=1}^n p_k \log_2\frac{1}{p_k}.
\label{shannon}
\end{equation}

The quantum equivalent to the R\'{e}nyi-$\alpha$ entropy is defined as
\begin{equation}
S_\alpha(\rho)=\frac{1}{1-\alpha}\log {\rm Tr} (\hat \rho^\alpha),
\end{equation}
where $\hat \rho$ corresponds to a density matrix of a quantum state. The von Neumann entropy, as the quantum analogue of Shannon entropy, is defined as the quantum R\'{e}nyi-$\alpha$ entropy in the limit $\alpha\rightarrow 1$ 
\cite{vNfunction}. 
The special case of $\alpha=2$ is of particular interest for several reasons, including its inherent connection to the purity of the state and its natural emergence from quantum phase-space sampling \cite{sampling}. For Gaussian states the 
R\'{e}nyi-2 entropy can simply be written as 
\begin{equation}
\mathcal{S}_2(\rho)=-\ln {\rm Tr} (\hat \rho^2)=\frac{1}{2}\ln(\det \gamma),
\end{equation}
that is, linked in a very straightforward way to the covariance matrix $\gamma$, which defines the corresponding Gaussian state and is measurable in an experiment.

The fact that the R\'{e}nyi-2 entropy arises naturally from phase-space sampling for Gaussian states provides a strong motivation to use this entropy to define entropic measures for Gaussian states. In general R\'{e}nyi-$\alpha$ entropies for $\alpha\neq1$ are not subadditive, thus quantities such as the quantum mutual information can become negative, and are then meaningless correlation measures. However, R\'{e}nyi-2 entropy satisfies a strong subadditivity inequality for all Gaussian states, the consequence of this is that it allows the core of quantum information theory to be consistently recast within the Gaussian regime, using the physically natural and simpler 
R\'{e}nyi-2 entropy as opposed to the von Neumann entropy (see \cite{adessoR2} for a detailed account of the correlations and information measures using R\'{e}nyi-2 entropy). 

\subsection{Gaussian states: classical or non-classical?}
In this section we would like  to address some fundamental aspects for bosonic bi-partite quantum systems (which include bi-partite Gaussian states). For such systems, there are two distinctly different notions of nonclassicality \cite{ferraro-paris,chille15,paris15}, the $P$-classicality and the $C$-classicality, the latter connected directly to quantum discord.

\subsubsection{The ${P}$-classicality}

The conventional nonclassicality criterion in quantum optics is related to the Glauber-Sudarshan $\mathcal{P}$-function 
\cite{Glauber_63}, one of the most used phase-space quasi-probability distributions as it diagonalizes the
density operator in terms of coherent states \cite{wigner}. Coherent states $\alpha$ are minimum uncertainty states
(cf. minimum uncertainty states of simple harmonic oscillator in quantum theory) which minimize the Heisenberg uncertainly relation and have equally distributed uncertainty between the conjugate variables. They are defined as eigenstates of the bosonic annihilation operator $\hat a$, $\left [ \hat a, \hat a^\dagger \right ] = 1$. The corresponding eigenvalue equation reads $\hat a \vert \alpha \rangle = \alpha \vert \alpha \rangle$, where $\alpha$ plays then the role of the coherent amplitude. Coherent states are the states with very well defined amplitude, only restricted by the minimum quantum uncertainty. Ideal laser light is described by quantum coherent state. The set of coherent states is overcomplete, i.~e., the states with different amplitudes are not orthogonal.

According to the $\mathcal{P}$-function 
criterion (also known as optical theorem \cite{wigner}), a state of a bi-partite system is classical, if its density matrix $\hat \rho$ can be represented as a statistical mixture of two-mode coherent states $|\alpha\rangle|\beta\rangle$  with well behaved $\mathcal{P}$-function, 
\begin{eqnarray}
\hat{\rho}=\int\int_{\mathbb{C}}\mathcal{P}(\alpha,\beta)|\alpha\rangle\langle\alpha|\otimes|\beta\rangle\langle\beta|\ d^2\alpha\  d^2\beta . \label{P-criterion}
\end{eqnarray} 
Here the  $\mathcal{P}$-function plays a role of classical probability distribution and uniquely determines the quantum state, there is a one-to-one correspondence between the density matrix and phase-space quasi probability distribution
$\mathcal{P}(\alpha,\beta)$. 

Note that in the nonclassicality criterion in quantum optics (\ref{P-criterion}), the density matrix is represented as an expansion over {\it non-orthogonal} basis states. And this is exactly the point where the discrepancy with the informational-theoretical criterion emerges and exactly that allows the $\mathcal{P}$-classical states 
to possess quantum correlations described by non-zero quantum discord, and thus be $C$-nonclassical.

\subsubsection{The $C$-classicality and inequivalence with the P-classicality}

As stated above, all non-product bi-partite Gaussian states
have been shown to have non-zero Gaussian discord \cite{adesso,Mista_14}
b\emph{}ut many of them are termed classical according to the conventional
nonclassicality criterion \cite{Glauber_63}, the $P$-classicality criterion described above.
Thus a wide range of states, normally perceived as classical,
exhibit according to the Gaussian discord quantum correlations
and should be classified as quantum. Recurring examples of
non-zero Gaussian discord in such seemingly classical states
raised doubts whether Gaussian discord is a legitimate measure.
This apparent discrepancy has first been discussed in \cite{ferraro-paris}:
the nonclassicality criteria can differ in the quantum-optical
realm and in information theory. Therefore states classified as
quantum in one context, can appear classical in the other. 

To specify the nature of correlations based on the different types of nonclassicality, let us first refer again to the definition of entanglement (\ref{ent}) formulated by Werner \cite{werner}. As discussed in \cite{ferraro-paris},
it has an immediate operational interpretation: separable, not entangled states can be prepared by local operations and classical communication between the two parties. This excludes a possibility of quantum character only at the first glance.  In information theoretical context, the entropic definition of discord (Definition II) shows that in quantum regime there is a mismatch in classically equivalent definitions of mutual information. The crucial point here is that if the definition (\ref{ent}) is written in terms of density matrices, this includes an expansion over genuinely non-orthogonal basis. The states $\hat \rho_{Ai}$, $\hat \rho_{Bi}$ maybe physically indistinguishable and therefore the locally available information about them may be incomplete. This is completely different from classical situation and is an example of quantumness in separable states captured by quantum discord 
\cite{ferraro-paris}. Thus from the information theoretical perspective there are different types of bi-partite separable states
\cite{classification}:\\ \\
1. {\it Bi-partite quantum states}:
\begin{eqnarray} 
\hat \rho_{AB} = \sum_{i} p_{i} \hat \rho_{Ai} \otimes \hat \rho_{Bi}
\label{QQ}
\end{eqnarray} 
allowing for both non-zero $A$-discord and non-zero $B$-discord. Remark: The expressions `$A$-discord', `$B$-discord' reflect the asymmetry involved in the definition
of quantum discord: the information about system $A$ is gained via measurement on $B$ or vice versa.\\ \\
2. {\it Quantum-classical (QC) states}:
\begin{eqnarray} 
\hat \rho_{AB} = \sum_{i} p_{i} \vert a_i \rangle \langle a_i \vert \otimes \hat \rho_{Bi},
\label{QC}
\end{eqnarray} 
where the states $\vert a_i \rangle $ form an orthonomal basis and the $\hat \rho_{Bi}$ are a set of generic non-orthogonal states. The states (\ref{QC}) have zero $A$-discord, but {\it non-zero} $B$-discord and cannot be cloned 
(broadcasted) locally. Local broadcasting means the procedure of locally distributing pre-established correlations in order to have more copies of the original state \cite{classification}.\\ \\
3. {\it Classical-classical (CC) states}:
\begin{eqnarray} 
\hat \rho_{AB} = \sum_{ij} p_{ij} \vert a_i \rangle \langle a_i \vert \otimes \vert b_j \rangle \langle b_j \vert ,
\label{CC}
\end{eqnarray} 
where the states $\vert a_i \rangle $ and $\vert b_j \rangle$ form an orthonomal basis. In case of CC-state, quantum state 
can be interpreted from the very beginning as a joint probability distribution that
describes the state of classical registers. That is, we can simply speak about
the embedding into the quantum formalism of a classical joint probability distribution.
Note that, as in the case of class 1 and class 2, this is not generally possible any more if at least 
one of the basis sets $\hat \rho_{i}$ correspond to generic non-orthogonal states. It is clearly seen that the Definition I of quantum discord (\ref{discord-modi}) is directly linked to this state classification. The CC-states are referred to as $C$-classical.

It has been shown in \cite{ferraro-paris}, that the sets of the $C$-classical and $P$-classical states are maximally inequivalent. There are many examples of the states that are classical according to one criterion and non-classical according to the other. The corresponding proofs are involved and out of the scope of this review, the interested reader is referred to \cite{ferraro-paris,classification}. The nonclassicality of the Quantum-Classical and Quantum-Quantum states is a key to quantum discord and ultimately, to a new research field of quantum coherence (see sec.~\ref{qu-coh} and, e.~g., \cite{coherence} and references therein).

\subsection{Operational meaning of quantum discord and entanglement activation from quantum correlations}

Quantum correlations beyond entanglement have found  applications in metrology and other quantum technologies \cite{adesso-review,streltsov-review,{lectures-gen-cor}}, giving quantum discord an operational meaning. Our understanding
and quantitative characterization of coherence as an operational resource can be facilitated by linking it to entanglement \cite{streltsov2015}. 
It has been first shown in \cite{adesso-noncl} that all non-classical correlations can be activated into entanglement, using auxillary system and C-NOT gates, giving correlations a new operational meaning in terms of resources for entanglement generation. Quantifying of nonclassicality can then be done by quantifying the resultant entanglement,
for which many tools for analysis are already known. For example, the relative entropy of quantumness,
which measures all nonclassical correlations among subsystems of a quantum system, is equivalent
to and can be operationally interpreted as the minimum distillable entanglement generated between the system
and local ancillae in the protocol of \cite{adesso-noncl}, which has been recently implemented in an experiment 
\cite{adesso14}. In a related work, it has been shown, that the quantumness of correlations as measured by the quantum discord is  related to the minimum entanglement generated between system and apparatus in a partial measurement process \cite{streltsov-ent}.

\subsubsection{Concept of entanglement distribution by separable ancilla}
Along the same lines, states with non-zero quantum discord can be used to share entanglement between distant parties without the need of an entangled carrier, as has been recently demonstrated in three independent experiments \cite{peuntinger13,Vollmer13,Fedrizzi13}. In 2003, Cubitt et al. \cite{cubitt} showed that mixed separable states can actually be used to distribute entanglement between two remote parties, which is counter-intuitive and impossible with pure separable states. Later this idea has been extended from qubits to continuous variables \cite{mista-distr}. For such Gaussian states the mechanism behind entanglement activation from initial mixed separable three-partite state can be unveiled in a lucid fashion and will be discussed in detail in this section. 
In all three protocols \cite{peuntinger13,Vollmer13,Fedrizzi13}, initially three systems $A$, $B$, and $C$ (single photons or light beams) are in pure states, and the global state is a product state with zero discord. Then a tailored dissipation is introduced to render a mixed but correlated state, with a particular correlation pattern.  At this stage two systems $A$ and $B$ and the ancilla $C$ are in a three-partite fully separable state with non-zero discord. Then $ C$ interferes first with $A$ and then with $B$. Upon this second interaction $A$ and $B$ become entangled. $C$ remains separable throughout the protocol. Let us stress, that the initial state $ABC$ is though separable but discordant, that is, all three modes are correlated in a particular fashion and the entanglement distribution by separable ancilla can be  interpreted as entanglement activation from quantum discord \cite{streltsov-distr,chuan-distr}.

\begin{center}
\begin{figure}[h]
\begin{center}
\includegraphics[width=10cm]{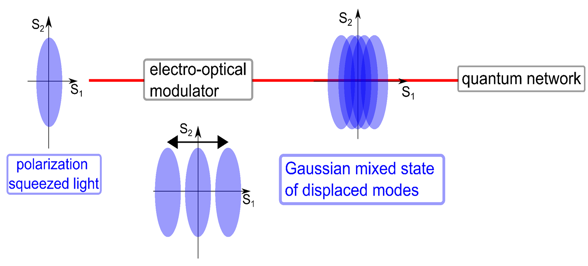} \end{center}
\caption{ (color online) Experimental preparation of the Gaussian mixed state from a pure input state (Credit: V. Chille). The initial state (here a pure squeezed state) is sent through an electro-optical modulator that displaces the squeezed state by some amount. There states are mixed together in post-processing to produce a Gaussian mixed state of the displaces modes.  }
\label{state-prep}
\end{figure}
\end{center}

\begin{center}
\begin{figure}[h]
\begin{center}
\includegraphics[width=\linewidth]{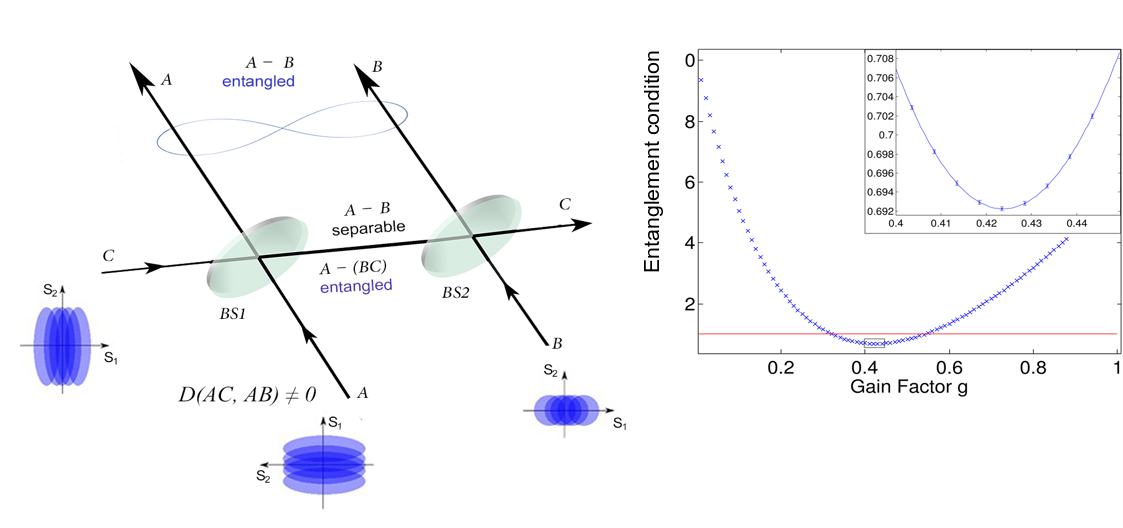} \end{center}
\caption{ (color online) Entanglement distribution by separable ancilla (see \cite{peuntinger13}). Right: Principle scheme. Initial state is a three-partite fully separable mixed state. BS: beam splitter. Shaded circles and ellipses: contours of the Wigner function, depicting the quantum states of the input modes, circles correspond to coherent and ellipses to squeezed states, respectively. Displayed, overlapping circles (elipses): random but correlated displacements in phase space (see Fig.~{\ref{state-prep}}). All the individual BS inputs are $P$-classical, but due to global correlations, entanglement can be extracted from discordant global state and localized between modes $A$ and $B$ . Left:  Entanglement distributed between modes $A$ and $B$. The experimental values for the 
criterion (\ref{Duan}) are depicted in dependence of the gain factor $g$. Due to the attenuation of the mode 
$B$ by 50\,\%, a gain factor about 0.5 yields a value smaller
than 1, i.e. below the limit for entanglement (red line). The inset zooms into the interesting
section around the minimum. The depicted estimated errors are so small because of the large amount of data taken. }
\label{ent-distr}
\end{figure}
\end{center}

\begin{center}
\begin{figure}[h]
\begin{center}
\includegraphics[width=\linewidth]{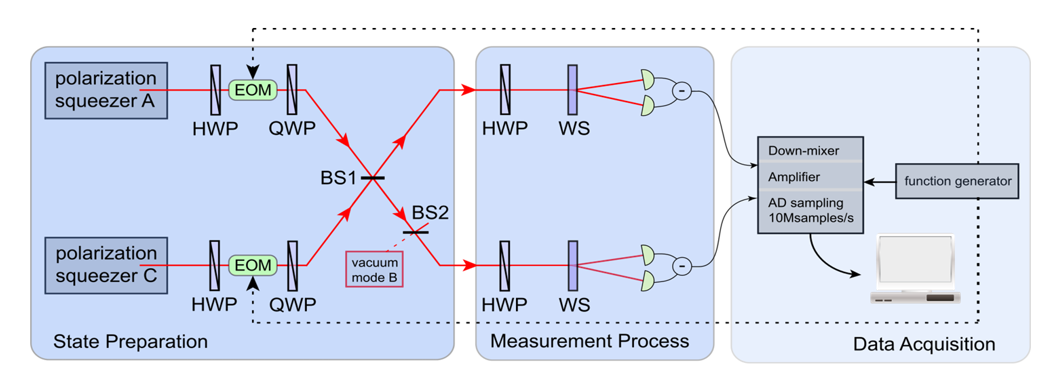} \end{center}
\caption{ (color online) Entanglement distribution by separable ancilla. Schematic of the experimental setup (taken from \cite{peuntinger13}). Used abbreviations: HWP: Half-wave plate; QWP: Quarter-wave plate; EOM: Electro-optical modulator; BS: Beam splitter; WS: Wollaston prism; (State Preparation) The polarization of two polarization squeezed states ($A$ and $C$) is modulated using EOMs and sinusoidal voltages from a function generator (dotted lines). The HWPs before the EOMs are used to adjust the direction of modulation to the squeezed Stokes variable, whereas the QWPs compensate for the stationary birefringence of the EOMs. The such prepared modes interfere with a relative phase of $\pi/2$ on a balanced beam splitter BS1 (also denoted BS1 in Fig.~\ref{ent-distr}). In the last step of the protocol, the mode $C$ interferes with the vacuum mode $B$ on a second balanced beam splitter BS2  (BS2 in Fig.~\ref{ent-distr}). (Measurement Process) A rotatable HWP, followed by a WS and a pair of detectors, from which the difference signal is taken, allows to measure all possible Stokes observables in the $\hat S_1$-$\hat S_2$-plane. To determine the two-mode covariance matrix $\gamma_{AB}$ all necessary combinations of Stokes observables are measured. Removing the second beam splitter of the state preparation allows us to measure the covariance matrix of the two-mode state $\hat\rho_{AC}$. (Data Acquisition) To achieve displacements of the modes in the $\hat S_1$-$\hat S_2$-plane we electronically mix the Stokes signals with a phase matched electrical local oscillator and sample them by an analog-to-digital converter.  }
\label{ent-distr-exp}
\end{figure}
\end{center}

\begin{center}
\begin{figure}[h]
\begin{center}
\includegraphics[width=12cm]{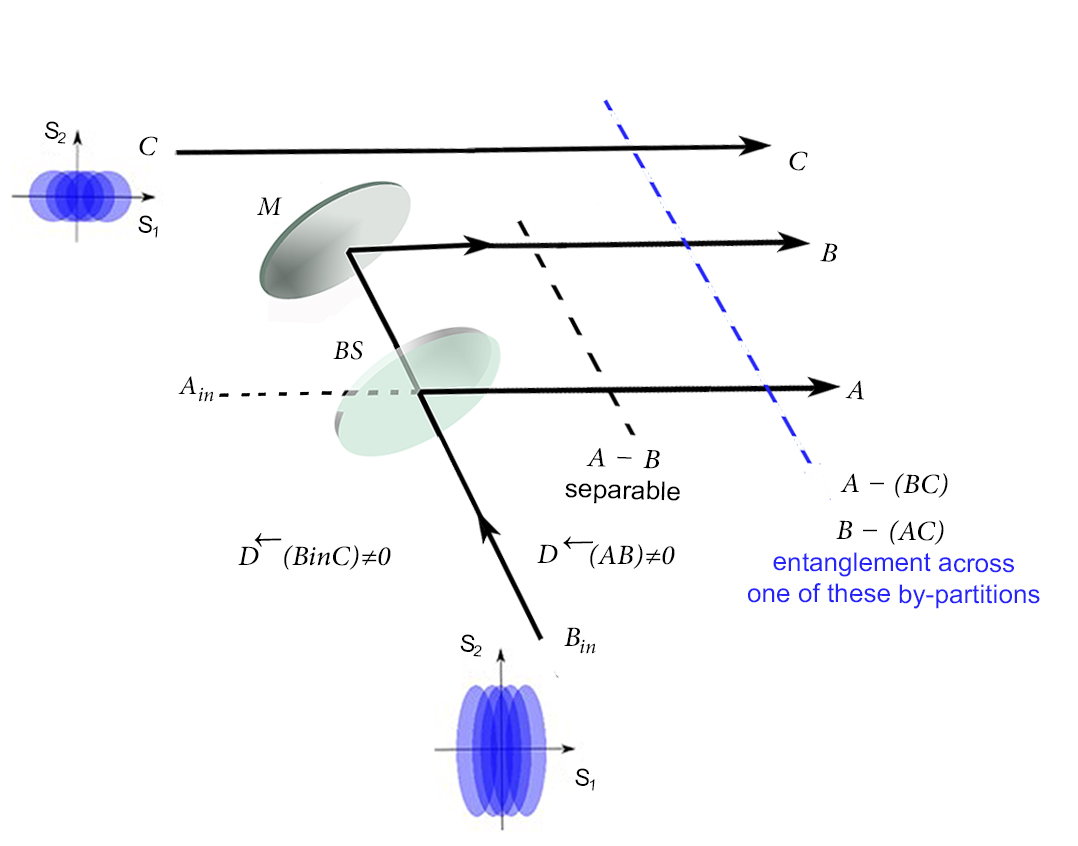} \end{center}
\caption{ (color online) Entanglement from discord using merely a passive operation. BS: beam splitter. Shaded displaced circles and ellipses: as in Fig.~\ref{ent-distr}. After the application of displacements, individual states of each mode $B_{in}$ and $C$ become highly mixed (and locally non-squeezed), but the total state $B_{in}C$ is correlated and has non-zero discord. The input mode  $A_{in}$ is in the vacuum state (unused BS port). The output three-mode state $ABC$ is entangled across the bi-partitions shown (see \cite{croal15}).   }
\label{entangBS}
\end{figure}
\end{center}

Let us revert to Gaussian states for a more detailed experimental illustration. Entanglement activation from discord has been demonstrated experimentally in different protocols with Gaussian states \cite{chille15,peuntinger13,Vollmer13,croal15}, where the crucial entangling operation was implemented as a beam splitter acting on a separable multi-mode state, which possesses discordant correlations.  A beam splitter (BS) is frequently used to generate entangled continuous variable states, if at least one of the inputs is a quantum squeezed state.  BS is passive and can only create entanglement if there is some quantumness initially. Several protocols  demonstrated experimentally that, remarkably, for mixed quantum states, a BS can create entanglement even from two input modes none of which exhibit any local squeezing, provided that they are correlated in a tailored way with a third one
(\cite{croal15} and references therein). Exactly this mechanism is behind the entanglement distribution by separable ancilla with Gaussian states \cite{peuntinger13,Vollmer13}.

\subsubsection{Quantum polarization variables and state preparation} \label{stokes}
First let us discuss the preparation of the initial fully separable three-partite mixed state with an example of a pure state squeezed in $x$-quadrature, $\hat x^{(0)}=e^{- r}$ (Fig.~\ref{state-prep}). Here $r$ is the
squeezing parameter and the superscript ``$(0)$'' denotes the
vacuum quadrature. In the particular implementation of  \cite{chille15,peuntinger13,croal15},  polarization variables described by Stokes operators (see e.g. \cite{Korolkova_02}) are used instead of quadratures:
\begin{eqnarray}
 \hat{S}_0=\left( \hat{a}^\dagger_x\hat{a}_x+\hat{a}^\dagger_y\hat{a}_y\right), \qquad \hat{S}_1=\left( \hat{a}^\dagger_x\hat{a}_x-\hat{a}^\dagger_y\hat{a}_y\right), \nonumber
\\ \hat{S}_2=\left(\hat{a}^\dagger_x\hat{a}_y+\hat{a}^\dagger_y\hat{a}_x\right),
\qquad \hat{S}_3=i\left(\hat{a}^\dagger_y\hat{a}_x-\hat{a}^\dagger_x\hat{a}_y\right) \label{stokes-op}
\end{eqnarray}
where $\hat{a}_{x,y}$ are the annihilation operators for photons linearly polarized in the $x$- and $y$- directions respectively. The $\hat{S}_0$ operator corresponds to the beam intensity whilst $\hat{S}_j$
describe the polarization state. The $\hat{S}_0$ operator commutes with the others, whereas the remaining operators obey the SU(2) Lie algebra, as indicated by the commutator: 
\begin{eqnarray}
\left [ \hat{S}_1, \hat{S}_2 \right ]= 2i \hat{S}_3, 
\end{eqnarray}
and the cyclics thereof. Stokes observables  $\hat{S}_j$ span the Poincare sphere, analogously to the Bloch sphere in the case of spin variables. Polarization squeezing is defined as quantum states with the uncertainty in one of the Stokes operators reduced below that of coherent polarization states \cite{Korolkova_02}. There are some subtleties in definitions of polarization squeezed states but they are not relevant for the current review as we turn to the strongly polarized case below and effectively use the quadrature-squeezed states.

Polarization description of light will become important also in the last section of this review, when speaking about classical entanglement in sec.~\ref{pol-cl-ent}. In this section, in quantum protocols with discordant correlation, the advantage of polarization squeezed states is merely practical: their measurement does not require the use of local oscillator, which is needed in standard homodyne detection and is associated with challenging requirements for phase-locking and high visibility of interference \cite{Korolkova_02}.  In contrast to this simply experimental benefit here, in sec.~\ref{pol-cl-ent} it will really be the specific polarization properties of vector fields and the Poincare sphere representation that render classical entanglement possible.  

We choose the state of polarization such that mean values of $\hat{S}_1$ and $\hat{S}_2$ equal zero while $\langle\hat{S}_3\rangle\gg0$. This configuration allows to identify the ``dark'' $\hat S_1$-$\hat S_2$-plane with the quadrature phase space. $\hat{S}_{\theta}, \hat S_{\theta + \pi/2}$ in this plane correspond to $\hat S_1, \hat S_2$ renormalized with respect to $\hat {S}_3 \approx S_3$ and can be associated with the effective quadratures $\hat x,\hat p$. A mode polarization-squeezed in $\hat S_1$ corresponds thus to $\hat x^{(0)}=e^{- r}$  and is shown in the leftmost part of Fig.~\ref{state-prep}. Then electro-optical modulator is used to add the noise in the form of random displacements to the squeezed observables (central part of Fig.~\ref{state-prep}). The actual creation of the Gaussian mixed state happens at the data acquisition stage, when the Stokes signals are electronically mixed with a phase matched electrical local oscillator and sampled by an analog-to-digital converter. Using further appropriate digital post processing, the Gaussian mixed state is prepared (the rightmost part of Fig.~\ref{state-prep}). Thus a thermal Gaussian state, a statistical mixture, in an individual mode is prepared from a quantum pure state. Discordant correlations are imposed on the initially product states at the stage of random displacements. The modulation patterns applied to different modes are chosen in a particular fashion (depending on the desired structure of correlations) such that random displacements in the modes are not independent.

\subsubsection{Experiment on entanglement distribution using separable states}
The protocol \cite{mista-distr} is depicted in Fig.~\ref{ent-distr} (left). 
Initially,  modes $A$ and $C$ are prepared in a momentum squeezed and position squeezed vacuum state,
respectively, with quadratures
$\hat x_{A,C}=e^{\pm r}\hat x_{A,C}^{(0)}$, $\hat p_{A,C}=e^{\mp
r}\hat p_{A,C}^{(0)}$, whereas mode $B$ is in a vacuum state with
quadratures $\hat x_{B}=\hat x_{B}^{(0)}$ and $\hat p_{B}=\hat p_{B}^{(0)}$. All the modes are then subjected to suitably
tailored local correlated displacements as described above and illustrated in Fig.~\ref{state-prep}:
\begin{eqnarray}\label{displacements}
\hat p_{A}&\rightarrow& \hat p_{A}- p,\quad \hat x_{C}\rightarrow
\hat x_{C}+ x,\nonumber\\
\hat x_{B}&\rightarrow& \hat x_{B}+\sqrt{2}x,\quad
\hat p_{B}\rightarrow \hat p_{B}+\sqrt{2} p.
\end{eqnarray}
The uncorrelated classical displacements $x$ and $p$ obey a
zero mean Gaussian distribution with the same variance
$(e^{2r}-1)/2$. The resultant state has been prepared by local operations and classical communication across $A-B-C$
splittings and therefore fully separable.

In the second step, modes $A$ and $C$ interfere on a balanced beam splitter (Fig.~\ref{ent-distr}, left). Upon this first interference, the state is separable with respect to $B-(AC)$ bi-partition and fulfils the positive partial transpose criterion with respect to mode $C$ and
hence is also separable across $C-(AB)$ \cite{werner}.

In the final step,  mode $C$ interferes with mode $B$ on another balanced beam splitter and this activates entanglement  between modes $A$ and $B$ verified by the sufficient condition for entanglement \cite{giovannetti-ent,Dong_07}: 
\begin{eqnarray}\label{Duan}
\Delta_{\rm norm}^{2}(g\hat x_{A}+\hat x_{B})\Delta_{\rm norm}^{2}(g\hat p_{A}-\hat p_{B})<1.
\end{eqnarray}
Here $g$ is a variable gain factor and variances $\Delta_{\rm norm}^{2}(.)$ are normalized  to  the  respective mean  values  of  $\langle \hat S_3 \rangle$ (the bright polarization component),  which
corresponds to the shot noise reference (see, e.g., \cite{Dong_07}). Note that the effective quadrature operators in Eq.~(\ref{displacements}) are the Stokes operators in the ``dark plane'' orthogonal to the bright component $\hat S_3$ renormalized  such that $\left [ \hat x_j, \hat p_j \right ]= i$,  as described above. This imposes the unit bound for the entanglement criterion and the weaker EPR criterion, which are essentially equivalent when
applied  to  observables with  trivial  commutation  rules \cite{giovannetti-ent}. With the appropriate recalibration and minimizing the left hand side of Eq.~(\ref{Duan}) with respect to $g$, the measurements depicted in Fig.~\ref{ent-distr} result in
\begin{eqnarray}\label{Duan2}
\frac{\Delta^{2}(g\hat x_{A}+\hat x_{B})\Delta^{2}(g\hat p_{A}-\hat p_{B})}{(g^2+1)^2}=0.8584 \pm 0.0005<1.
\end{eqnarray}
for a gain of $0.4660\pm0.001$. We get fulfilment of the criterion for any $r>0$, which confirms successful entanglement distribution (see Fig.~\ref{ent-distr}, right). 
 
The experimental realization is divided in three steps: state preparation, measurement, 
and data processing. The corresponding setup is depicted in Fig.~\ref{ent-distr-exp}. 
The states involved are Gaussian quantum states and, as mentioned earlier in this section,
are completely characterized by their first moments and the covariance matrix $\gamma$ comprising all second moments.
To ensure the separability of mode $C$, the correlations between mode $A$ and $C$ after BS1 has been evaluated. For that, multiple pairs of Stokes observables ($\hat S_{A,\theta},\hat S_{C,\theta}$) has been measured, $\theta$ is the angle in the $\hat S_1$-$\hat S_2$-plane between $\hat S_{0^{\circ}}$ and $\hat S_{\theta}$. In this way, the experimentally measured
 covariance matrix $\gamma_{AC}$ has been obtained and separability of the state has been verified \cite{peuntinger13}. The output covariance matrix $\gamma_{AB}$ has been measured after BS2 and Bob verifies that the product entanglement criterion~(\ref{Duan}) is fulfilled as illustrated in Fig.~\ref{ent-distr}, right.
That proves the emergence of entanglement.
The used gain factor $g$ considers the slightly different detector
response and some intentional loss  at Bob's beam
splitter. The clearest confirmation of entanglement can be seen for $g_{\rm opt} = 0.4235\pm0.0005$ 
(Fig.~\ref{ent-distr}, right). With an appropriate renormalization it corresponds to the value of the product entanglement criterion~(\ref{Duan}) of $0.8584\pm0.0005<1$ which verifies the successfull entanglement distribution. This is the only step of the protocol, where entanglement emerges, thus demonstrating the remarkable possibility to entangle the remote parties Alice and Bob by sending solely a separable auxiliary mode $C$.

\subsubsection{Entanglement from discord}
The performance of the protocol can be explained using the structure of
the displacements (\ref{displacements}). Entanglement distribution without sending entanglement highlights vividly the important role played by classical information in quantum information protocols. Classical information lies in our knowledge about all the correlated displacement involved.  This allows the communicating parties to adjust the displacements locally to recover through clever noise addition quantum resources initially present in the input quantum squeezed states. Mode $C$
transmitted from Alice to Bob carries on top of the sub-shot noise quadrature of the input squeezed state the displacement noise which
is anticorrelated with the displacement noise of mode $B$. Therefore,
when the modes are interfered on the second BS,
this noise partially cancels out in the output mode $B$ when
the light quadratures of both modes add. Moreover, the residual noise in 
position (momentum) quadrature in $B$ is correlated (anticorrelated) with
the displacement noise in position (momentum) quadrature in mode $A$ after the first BS, again initially squeezed.
Due to this the product of variances in criterion (\ref{Duan}) drop below the value for separable states and thus entanglement between  modes $A$ and $B$ emerges.

Similarly, for the discrete-variable experiment of \cite{Fedrizzi13}, initially $A$ and $B$ represent an entangled pair of photons, which are shared between Alice and Bob. In analogy with introducing globally correlated displacements in Gaussian, continuous-variable case, this entanglement is destroyed by randomly mixing the four different types of possible entangled states, the Bell states (\ref{Bell}). This procedure effectively prepares a separable mixed state between photons $A$ and $B$ of a tailored form, carrying distinct correlations. The information carrier, photon $C$
is similarly prepared in a specific mixed state. The required quantum interference between $A$ and $C$ and then $B$ and $C$ is accomplished by passing the photons through a quantum gate. The equivalent Gaussian operation in \cite{peuntinger13,Vollmer13} is performed by letting the corresponding optical modes interfere on a beamsplitter.

Altogether, both in discrete and in continuous variables cases,  this procedure produces a very particular three-party state, which is specifically tailored for the needs of the experiment. It has remarkable separability properties. 
The presence of correlated noise results in non-zero discord at all stages of the protocol. The role of discord in entanglement distribution has been recently discussed theoretically~\cite{streltsov-distr,chuan-distr}. The requirements devised there are reflected in the particular separability properties of the global state after the interaction of modes $A$ and $C$ on the first BS. Upon interaction of $A$ and $C$ on this BS, the state $\hat \rho_{A B C}$  contains discord and entanglement across $A-(BC)$ splitting and is separable and discordant across $C-(AB)$ splitting as required by the protocol (Fig.~\ref{ent-distr}, left). Thus the resultant entanglement bewteen $A$ and $B$ can be seen as activation of the bound entanglement of $A-(BC)$ or, ultimately, as entanglement activation of the initial discord between the three input modes.

The same mechanism allows for generation of a three-partite entangled state by splitting on a BS a thermal state correlated with a vacuum mode \cite{croal15} (Fig.~\ref{entangBS}). Here the BS generates entanglement from two input modes $A_{in},B_{in}$, where $A_{in}$ is a vacuum mode, i. e., merely an empty input port of the BS. $B_{in}$ is in a thermal state.  None of $A_{in},B_{in}$ exhibit any local squeezing, but they (or here actually only $B_{in}$) are correlated in a tailored way with a third mode $C$. In this protocol the created entanglement does not occur between the output modes of the BS but instead it emerges between one output mode and the remaining two modes taken together (see Fig.~\ref{entangBS}). This phenomenon is a key element of the protocols for entanglement distribution with separable states above  
\cite{peuntinger13},  for entanglement sharing \cite{Mista_13} and others. 
That is, there are fully separable and only globally non-classical three-mode states, possessing discordant correlations,
that can lead to entanglement using a beam splitter. A similar
effect happens also in the qubit case, where the CNOT gate can
generate entanglement by acting on a part of a suitable
three-qubit fully separable state, whereas it leaves the output of
the operation separable \cite{cubitt}. The local state may
appear unsuitable as a quantum resource, being $P$-classical (but not $C$-classical!). However, when being a
part of a larger correlated state, it can become a source of tailored entanglement. 

Entanglement generated from discord (or ultimately, from quantum coherence) is between different modes of light. However, there exist another type of entanglement where quantum correlations arise within a single optical mode or optical beam, between its different degrees of freedom. Such entanglement, often referred to as classical entanglement, is this subject of the next section.

\section{Classical entanglement in optical fields: intra-system quantum correlations }\label{pol-cl-ent}

The concept of { classical vs quantum and local vs non-local} entanglement have been first discussed in late 90ies, in the context of pioneering achievements in generation of quantum superposition states of atoms. A vivid example of {locally} entangled state is a mesoscopic Schor\"odinger cat - like
state of cold atoms realized with trapped Be ions in 1996 in the group of Wineland \cite{wineland}. They created the following state of a single $^9{\rm Be}^+$ laser-cooled ion:
\begin{equation}
\vert \psi \rangle= \frac{1}{\sqrt{2}} \Big ( \vert x_1 \rangle \vert \uparrow \rangle + \vert x_2 \rangle \vert \downarrow \rangle \Big ) \label{atom}
\end{equation}
and interpreted it as a Schr\"odinger-like state of two spatially separated coherent harmonic oscillator states, $\vert x_{j} \rangle$, $j=1,2$. Here $\vert x_{j} \rangle$ denote localized wave-packet states corresponding to two spatial positions of the atom; $ \vert \uparrow \rangle,  \vert \downarrow \rangle$ are two distinct internal electronic quantum states of the atom (hyperfine ground states). The state of Eq.~(\ref{atom}) can be seen as entangled: this state is nonseparable in the sense that it cannot be written as the product of two kets, formally thus obeying the definition of entanglement for general states formulated in Eq.~(\ref{ent}). However, it is not non-local, as states $\vert x_{j} \rangle$ and $ \vert \uparrow \rangle,  \vert \downarrow \rangle$ refer to two different degrees of freedom of one and the same object. The state (\ref{atom}) is hence prototypic for ``local entanglement'' of states which cannot be separated spatially.  Alternatively and maybe more specific, one could say that non-local entanglement refers to states on which one can perform two independent measurements, the outcome of which each show a statistical distribution, but which, when compared to each other reveal strong correlations. In this sense local entanglement refers to states for which two such measurements are not possible. {Initially, this distinct feature of  some type of entanglement being local  (as opposed to non-local entanglement like in (\ref{Bell})) has been perceived as a signature of classical entanglement \cite{spreeuw}. However, this is not generally true. We should rather speak of three types of entanglement}:
\begin{itemize}
\item {{\it nonlocal} {\it quantum entanglement}}: entanglement between separate entities (particles, optical beams, atomic ensembles etc); example - two two-level entangled systems in one of the states in Eqs.~(\ref{Bell});
\item {{\it intra-mode} or {\it local} {\it quantum entanglement}: quantum entanglement between different properties of a single entity; example - a single atom in state (\ref{atom})} . 
\item {{\it intra-mode} {\it classical entanglement}: entanglement between different properties of a single entity; examples will be discussed below.}
\end{itemize}
{In all three cases, the states} generically can have the form of the Bell states (\ref{Bell}), but the analogy goes only as far - classical entanglement, for example,  cannot be used to demonstrate EPR paradox. However, it has its own uses, as a reliable mathematical tool  \cite{eberly2011,saleh2012}, for metrology applications \cite{leuchs2014,leuchs2015} and others.{ We will return to the distinction between classical and quantum intra-system entanglement at the end of sec.~\ref{hilbert}.} 

The formal equivalence between the state of the spin-$\frac{1}{2}$ particle on the Bloch sphere and the polarization state of light on the Poincare sphere  (as well as any other two level systems) allows an easy transfer of the concept of classical entanglement of atomic degrees of freedom to polarization properties of electromagnetic fields and optical beams \cite{spreeuw}, as was later picked up and applied to study the topological phase structures associated with polarization and spatial mode transformations of optical vortex beams \cite{souza2007,souza2014} and to a number of open question in quantum polarization theory \cite{eberly2011,eberly2016,luisclassent}. Such intra-mode entanglement in optical vector fields will be the main focus of this review. For recent overviews on the topic see \cite{ghose,aiello2015}.

The term itself, {\it classical entanglement}, is in a way controversial and can be seen as an oxymoron raising some critics \cite{boyd}. We will use this term  as it has established itself historically. We need to keep in mind that the use of classically entangled light in, e.g., high-precision measurements or even in quantum information is very specific and different from uses of inter-mode quantum entanglement. We understand classical entanglement as particular nonseparable structures in optical fields, its equivalence to the conventional quantum entanglement occurs partially at a  formal level. It can not necessarily  be used to address particular fundamental questions in quantum mechanics. The concept of coherence which underlined all the phenomena discussed so far in this review, is also crucial here, as classical entanglement is clearly a manifestation of certain coherence properties in vector fields. 

\subsection{Polarization in optics, product vector spaces and optical cebits} \label{hilbert}
In recent years, multi-facet applications of optical beams have led to generation of light fields with non-trivial geometries, for example, radially polarized beams, doughnut-shaped beams, Laguerre-Gauss beams, tightly focused beams. Highly non-paraxial fields and 3D light fields  depart from traditional idea of an optical beam with a given direction of propagation, a specific transverse plane and a cylindrical symmetry. For such fields, definitions of degree of polarization, degree of optical coherence and other aspects of polarization theory should be reconsidered \cite{eberly2011,eberly2016,luisclassent}, as is also the case with quantum degree of polarization etc in quantum optics. Formal use of entanglement theory for classically entangled optical vector fields played important role  in answering these questions. 
 
The analogy starts with introducing the Hilbert space of  polarization vectors \cite{spreeuw,eberly2011,ghose}. Like the Hilbert space in quantum mechanics, the classical Hilbert space is spanned by state vectors, in this case basis vectors describing the polarization states or other degrees of freedom. In case of polarization, the counterparts of bra- and ket-vectors are the column and row vectors corresponding to Jones (or, alternatively, Stokes) polarization vectors on the Poincare sphere (see also sec.~\ref{stokes} and text around Eq.~(\ref{stokes-op})). When we speak about separability and entanglement in quantum mechanics, it presupposes the existence of the product Hilbert space (as opposed to the direct sum). When  we think of Eq.~(\ref{Bell}), the Bell states are defined on the Hilbert space formed by the product of individual subspaces $A$ and $B$. For polarization states, it can be created by combining polarization and spatial degrees of freedom. Following \cite{spreeuw}, we can introduce some basis for the polarization subspace, say $\{ \vert V ), \vert H )\}$ for vertically $V$ and horizontally $H$ polarized optical modes  or $x,y$ polarizations in Eq.~(\ref{stokes-op}) (see also Fig.~\ref{poincare}a). This defines a {\it polarization-cebit}. Here {\it cebit} is the classical counterpart of qubit, a two-level system which state is a linear combination of the given basis vectors $\{ \vert V ), \vert H )\}$ (in the same spirit, we can  also introduce $c$-spin, if reverting to spin degrees of freedom, $ \vert \uparrow ),  \vert \downarrow )$). Similarly, a subspace of beam position can be introduced, $\{ \vert x_1 ), \vert x_2)\}$, and a {\it position-cebit} defined \cite{spreeuw}. The elements of the product space spanned by $\{ \vert x_1 V ), \vert x_1 H ), \vert x_2 V ), \vert x_2 H )\}$ are then four-vectors, $\vert \Psi )$:
\begin{eqnarray} \label{4vector}
\vert \Psi ) &=&\left (
\begin{array}{c}
	\varepsilon_{x_1 V} \\ \varepsilon_{x_1 H} \\ \varepsilon_{x_2 V} \\\varepsilon_{x_2 H}
\end{array} \right ) \\ &=& \varepsilon_{x_1 V} \vert x_1 V ) + \varepsilon_{x_1 H} \vert x_1 H ) + \varepsilon_{x_2 V} \vert x_2 V ) + \varepsilon_{x_2 H} \vert x_2 H ) \nonumber
\end{eqnarray}
where $\varepsilon_{x_j Y}$ is the electromagnetic field amplitude of the $Y$ polarization component of the beam at the position $x_j$.
 
Now consider a pair of beams with equal intensity and orthogonal polarization:
\begin{eqnarray}
\vert \Psi ) = \frac{1}{\sqrt{2}} \Big ( \vert x_1 V ) + \vert x_2 H )\Big ),\label{class-Bell}
\end{eqnarray}
 and note, we consider this pair of beams, $x_1$ and $x_2$, {\it as a single object}. As a single entity, this beam pair in neither in a single, pure polarization state, nor in a single, pure position state. If we perform measurements on this object, we will find that the mesurement outcomes are correlated but they are not statistical as discussed above below Eq.~(\ref{atom}). Let us attempt to measure its polarization using rotatable polarizer. If we rotate the polarizer so that it transmits horisontal polarization (we measure $\vert H )$ state of the polarization-cebit), we will find that it transmits only the beam at position $\vert x_2)$ (we measure the position cebit in state $\vert x_2)$). In this sense we can say that   $\vert H )$ polarization correlates with $\vert x_2)$ position and $\vert V )$ polarization correlates with $\vert x_1)$ position.  Note, that this is not a correlation between different measurements of light excitation, but rather a correlation between filter orientations, or between a filter orientation and a measurement with a detector. This is exactly the same correlation pattern as in Eq.~(\ref{Bell}). In this sense we can say that the state of Eq.~(\ref{class-Bell}) is entangled. As in Eq.~(\ref{Bell}) the  state of the qubit $A$ is entangled with the state of the qubit $B$, for Eq.~(\ref{class-Bell}) we can say that the polarization-cebit of the beam pair is entangled with its position-cebit \cite{spreeuw}.

Eberly et al \cite{eberly2016} generalized the notion of classical entanglement even further, advocating an inherent link between quantum and classical optics, uniting them via common elementary notions of interference, polarization, coherence, complimentarity and entanglement. In their view, the association of entanglement with quantum theory is unnecessary and there is no need in identifying two different types of entanglement as we, following \cite{spreeuw} and others, did above. Eberly et al \cite{eberly2016} take a view that ``entanglement is a vector space property, present in any theory with a vector-space framework'' and there is no distinction between the quantum and classical entanglement. According to them, ``definition of entanglement is simply nonseparability of sums of product states that exist in different vector spaces'' \cite{eberly2016}.

The nature of the phenomenon once coined {\it classical entanglement} is still being widely discussed and open questions arise. Whereas the vector-space approach seems to hold for all the states considered to be classically entangled and doesn't seem to lead to any controversies, implications of this vector space property for different physical systems are not uniform.  Erwin Schr\"odinger introduced the theory we now know as quantum mechanics as wave mechanics. The language of wave optics is innately capable of capturing features inherent to quantum systems, for example, coherence, interference, superposition, etc. This link is lucidly discussed in \cite{eberly2016} and their treatment easily incorporates on the same footing classical optics and a range of quantum - or merely non-trivial? - properties of light. Nevertheless, the main open and controversial question in this context remains which, if any at all, aspects of all these phenomena are genuinely quantum. For instance, the states of type Eq.~(\ref{class-Bell}) admit fully classical description and we can say that the classical entanglement manifests itself entirely in intra-system correlations, in nonseparable state of different degrees of freedom of a single system. In contrast, there is no immediate classical equivalent to (\ref{atom}), which is a quantum superposition state of a cold atom, and is easier to accept as ``locally entangled''. {The differences in properties of states (\ref{atom}) and (\ref{class-Bell}) lends itself naturally to highlight the distinction between quantum and classical locally entangled states. Both states exhibit nonseparability of superpositions of product states that exist in different vector spaces. }

{ One of the profound differences that come into play here is that the joint Hilbert space (which structure then exhibits nonseparability) is spanned by the {\it mode functions}, whereas the measurement results are determined by the {\it excitations} ``living'' in this space. These excitations can be described using different bases of mode functions. A measurement is always done involving the excitation degree of freedom. In case of mode functions, we rather speak of filtering, the simplest example is the separation of different polarization modes using polarizing beamsplitter (PBS). The impact of this difference can be seen better 
using the defined earlier four mode basis (\ref{4vector}). Let us denote this state as $\vert k_{1V}, l_{1H}, m_{2V}, n_{2H} \rangle= \vert k, l, m, n\rangle$. Now let us keep the same mode function structure $\vert k, l, m, n\rangle$, that is the same structure of the Hilbert space, but consider  different types of excitation.  First consider a single photon, a single excitation in a mode denoted $\vert 1 \rangle$ with no excitation denoted $\vert 0 \rangle$. We start with $\vert k, l, m, n\rangle=\vert 0, 1, 0, 0\rangle=\hat a^\dagger_{1H}\vert 0, 0, 0, 0\rangle$, where $\hat a^\dagger_{1H}$ is a creation operator acting on the spatial mode $1$ horizontally polarized and describes a sigle excitation in this mode. After filtering on the PBS oriented under $45^\circ$, the state is transformed into the diagonal polarization basis, \{$\vert D \rangle, \vert A \rangle\}$:}
$$
\frac{\hat a^\dagger_{1D}+\hat a^\dagger_{1A}}{\sqrt{2}} \, \vert k_{1A}, l_{1D}, m_{2V}, n_{2H} \rangle =
\frac{\vert 0, 1, 0, 0\rangle + \vert 1, 0, 0, 0\rangle}{\sqrt{2}}.
$$
{This state is the quantum entangled  state of type $\vert 01 \rangle + \vert 10 \rangle$, a strict correlation of one photon in one arm and no photon in the other or vice versa \cite{bjork}. However if we excite a coherent state $\vert \alpha \rangle$ in the same mode $\vert x_1 H )$ and do the same filtering, we arrive at the state} 
$$
{\rm e}^{-\vert \alpha \vert^2 /2}{\rm e}^{\alpha \hat a^\dagger_{1H}} \, \vert 0, 0, 0, 0 \rangle = 
{\rm e}^{-\vert \alpha \vert^2 /2} \exp \left(\frac{\alpha}{\sqrt{2}} \hat a^\dagger_{1D} + \frac{\alpha}{\sqrt{2}} \hat a^\dagger_{1A}\right ) \, \vert 0, 0, 0, 0 \rangle = \vert \frac{\alpha}{\sqrt{2}}, \frac{\alpha}{\sqrt{2}}, 0, 0 \rangle
$$
{which does not exhibit any correlations in the new basis, because the projection noise of the measurements of the coherent states in the two modes is statistically independent. If we allow the state to fluctuate in time, $\alpha=\alpha(t)$, the two modes will show classical correlations in addition to the underlying quantum projection noise but will not exhibit classical entanglement. If we, however, excite a coherent state $\alpha$ jointly in modes $1H$ and $2V$, the output state reads:
$$
{\rm e}^{-\vert \alpha \vert^2 /2} \exp \left(\frac{\alpha}{\sqrt{2}} \hat a^\dagger_{1H} + \frac{\alpha}{\sqrt{2}} \hat a^\dagger_{2V}\right ) \, \vert 0, 0, 0, 0 \rangle = \vert 0, \frac{\alpha}{\sqrt{2}}, \frac{\alpha}{\sqrt{2}}, 0 \rangle.
$$
Obviously, this again does not look like a quantum entangled state and indeed measurements on the excitations of the modes will not reveal any quantum entanglement. However, if one filters in the appropriate degrees of freedom and measures the field excitation for the different filter positions, then one will find correlations between the different filter orientations. This is what is referred to as classical entanglement.} 
{The more detailed formal analysis of this characteristic difference between local entanglement quantum and classical is beyond the scope of this review and requires further investigation.} In what follows, we take a view of classically entangled systems as of those that possess certain nonseparability properties and consequently exhibit intra-system correlations which can enhance their performance beyond the established classical thresholds, e.~g., in metrology or in quantum information processing. {The optical states discussed below do not fall under {\it local quantum entanglement.}}

\subsection{Optical coherence and degree of polarization}\label{coh-pol}

In optics, nearly in every setting, several degrees of freedom are involved. This and indeed, as featured in \cite{eberly2016}, interference, polarization, coherence, complimentarity and entanglement of the vectorial optical fields are key notions in this general framework, by far not yet fully explored. In this section we briefly review the inner links between these notions which is very illuminating for understanding the richness and versatility of optics, classical and quantum.

In the  Hilbert space constructed in sec.~\ref{hilbert} we can define ${\bf \Gamma}=\vert \Psi )( \Psi \vert$ with $\vert \Psi )$ given in Eq.~(\ref{4vector}). ${\bf \Gamma}$ is Hermitian and non-negative and, extending quantum-classical analogy, it can be associated with a quantum density matrix $\rho= {\bf \Gamma}/\tr{\bf \Gamma}$ \cite{luisclassent}. We can also define the corresponding subsystems, 
${\bf \Gamma}_p$ for polarization (${\bf \Gamma}_s$ for spatial), by removing the spatial (the polarization) degrees of freedom. Following Luis \cite{luisclassent} and Berry and Sanders \cite{berry}, we define the degree of classical entanglement by linear entropy:
\begin{eqnarray}\label{cl-ent-degr}
{\cal E}^2 = 2 \left [ 1-\frac{\tr ({\bf \Gamma_j^2})}{(\tr{\bf \Gamma_j})^2}\right ] , \qquad j=s,p.
\end{eqnarray}
Maximal classical entanglement ${\cal E}=1$ occures for pure nonseparable states such as Eq.~(\ref{atom}), (\ref{class-Bell}). Minimum ${\cal E}=0$ 
corresponds to pure completely factorizable (fully separable) states. 

Coherence in vector fields can be defined and quantified differently, depending on which aspects are relevant to the problem in question. Let us now consider different ways to access coherence and their connection to degree of classical entanglement. The global amount of coherence $\mu_g$ present both in spatial and polarization degrees of freedom can be defined as Hilbert-Schmidt distance between ${\bf \Gamma}$ and $4 \times 4$ identity matrix representing fully incoherent and fully unpolarized light \cite{luisclassent}. It can be expressed via the linear entropy \cite{berry,wolf} using the trace of coherence or density matrix, or matrix ${\bf \Gamma}$, as appropriate. Global coherence $\mu_g=1$ is maximal for the pure state 
$\tr({\bf \Gamma}^2)=(\tr{\bf \Gamma})^2$. Minimum coherence $\mu_g=0$ corresponds to fully factorizable states. At the first glance, looking at the  invariance properties of $\mu_g$, there is no definite relation between classical entanglement and global coherence but the intrinsic link present here can be invoked using some formal equivalences to quantum information theory as we show at the end of this subsection. 

However, there is a definite relation between trace coherence $\mu_t$ or spatial coherence $\mu_s$ and classical entanglement. These correspond to the degree of coherence for vector electromagnetic fields defined in \cite{wolf},
\begin{eqnarray}
\mu_s = \mu_t = \frac{\langle {\bf E_2^\dagger}{\bf E_1}\rangle}{\sqrt{I_1I_2}}, \qquad {\bf E_j} =\left (
\begin{array}{c}
	\varepsilon_{x_j H}\\\varepsilon_{x_j V}
\end{array}\right ), \qquad I_j = \langle \vert {\bf E_j} \vert^2 \rangle,
\end{eqnarray} 
and can be also expressed in terms of $\tr{\bf \Gamma}$, hence the name trace coherence (for details see \cite{luisclassent,wolf}). Introducing predictabilities $\delta_j, j=s,p$ of the location and polarization states,
\begin{eqnarray}
\delta_s = \frac{I_1-I_2}{I_1+I_2}, \qquad \delta_p = \frac{I_x-I_y}{I_x+I_y},
\end{eqnarray}
one can verify that classical entanglement amounts to the product of unpredictability and incoherence:
\begin{eqnarray} \label{ent-coh}
{\cal E}^2 = \left ( 1-\delta_j^2 \right ) \left ( 1-\vert \mu_j \vert^2 \right ), \quad j=s,p.
\end{eqnarray}
For given predictability, larger classical entanglement means larger incoherence $1-\vert \mu_{j}\vert$. In contrast to quantum entanglement and quantum inter-system correlations, for which higher correlations mean higher coherence, larger classical entanglement and intra-system correlations mean lesser coherence. Is that surprising?

To answer this question, consider the link between classical entanglement, coherence and degree of polarization. It can be found looking at maximal spatial and polarization coherences, $\vert \mu_j\vert$. For polarization subspace, $\vert \mu_p \vert$ corresponds to the standard degree of polarization $P$ defined in terms of ${\bf \Gamma_p}$ and incidentally, equals also to the maximal spatial coherence, $\vert \mu_s \vert$:
\begin{eqnarray}
P^2=\vert \mu_p \vert^2 = \vert \mu_s \vert^2 = 2 \frac{\tr ({\bf \Gamma_j^2})}{(\tr{\bf \Gamma_j})^2} - 1, \quad j=s,p.
\end{eqnarray}
Using Eq.~(\ref{cl-ent-degr}) and re-arranging, we obtain a lucid connection between polarization, coherence and classical entanglement:
\begin{eqnarray}
P=\frac{\vert {\bf S_1} + {\bf S_2}\vert}{I_1 + I_2}=\vert \mu_j \vert = \sqrt{1-{\cal E}^2}, \quad j=s,p,
\end{eqnarray}
where ${\bf S_k}$ are Stokes vectors at positions $ k=1,2$. Their components are Stokes parameters $S_{1,2,3}$, 
classical counterparts of Eq.~(\ref{stokes-op}) and $I_j$ are light intensities at positions $ k=1,2$, $I=S_0^2$. As first discussed in \cite{luisclassent}, this bring us to conclusion that classical entanglement as measured by $\cal E$ and degree of polarization (or degree of spatial coherence of vector fields) are complementary features. Perfectly polarized light has to have zero degree of entanglement, as this would mean a pure polarization (spatial) state, thus the state of polarization (position) cebit is completely  defined and hence the state can be fully factorized (see also end of sec.~\ref{hilbert} around Eq.~(\ref{class-Bell})).  

Qian and Eberly \cite{eberly2011}  studied the connection between polarization and classical entanglement in the framework of entanglement theory in quantum information. They have re-expressed the degree of polarization using Schmidt theorem and shown that the Schmidt decomposition automatically delivers a useful weight parameter $K$ which counts the noninteger effective number of dimensions needed by the optical field:
\begin{eqnarray}
P^2=1-2\left ( 1-\frac{1}{K} \right ).
\end{eqnarray}
The Schmidt weight $K$ varies from $1$ to $3$ on the unit polarization sphere, $K=1$ being completely polarized and $K=3$ completely unpolarized fields. In the same spirit as above, $K=3$ is maximal entanglement. Intermediate values of $K$ represent intermediate degrees of entanglement (partially polarized light) \cite{eberly2011}. The degree of polarization of an optical field then corresponds to the degree of separability of the two disjoint Hilbert spaces corresponding to the polarization and spatial degrees of freedom. The relation between degree of entanglement and degree of polarization allowed Eberly et al \cite{eberly2016} to develop an analog of complimentarity showing that
\begin{eqnarray}
P^2 + C^2 =1,
\end{eqnarray}
where $C$ is degree of entanglement in two degrees of freedom measured by concurrence \cite{wootters}. In the same work, they analyze in detail coherence properties of optical fields, their connection to purity, polarization, and classical entanglement. This discussion is closely linked to the degree of polarization of higher order \cite{klimov}. There it is argued that in order to fully characterize the polarization one needs measure the polarization degree to all orders.


An interesting example of linking entanglement, non-locality and optical coherence from a different perspective is given in the work from Saleh group \cite{saleh2012}. There they show that Bell's measure of non-locality (e. g., Clauser-Horne-Shimony-Holt (CHSH) inequality) performed as polarization and spatial parity analysis of electromagnetic fields, can be used as a more precise measure of classical optical coherence. The optical coherence used in their work is the overall beam coherence defined using the linear entropy and corresponds to $\mu_g$ discussed above. The extended measure allows to clearly differentiate between incoherence associated with statistical fluctuations (e.g. partial coherence due to fluctuations of the source or due to propagation in a random medium) and incoherence in a beam connected to ignoring some of its degrees of freedom, classically entangled with the observed ones. Such extended notion of optical coherence is indispensable in cases where multiple degrees of freedom are relevant, for example, when a double slit experiment is performed with vector fields. As experimentally confirmed in \cite{saleh2012}, the ``classical'', CHSH-type Bell-measure identifies uncertainty present in each degree of freedom as a result of the classical entanglement between different degrees of freedom. The experimental results are distinct from any uncertainty originating from statistical fluctuations. The mathematical tools used have been directly mapped from quantum information theory using the formal equivalence between nonseparability of classically entangled and of quantum entangled states.

\subsection{Classical entanglement as a resource: quantum information and emerging technologies}\label{resource}

Looking at polarization and entanglement properties of light, we thus can say that  only homogeneously polarized light fields are fully separable. Thermal light is necessarily classically entangled. Moreover, an ideal thermal light field is a Bell state that violates ``local'' Bell inequality between different degrees of freedom (\cite{eberly2016,ghose,goldin,borges} and references therein). This triggered research on uses of classical light for information processing. In particular, Spreeuw has discussed in detail quantum information processing based on classical wave optics using the structural nonseparability of vector fields \cite{spreeuw2001}. Unitary operations with single and multiple cebits, one- and two-cebit logic gates, GHZ-states, error correction codes, teleportation protocols and other elements of quantum information networks can be introduced \cite{ghose,spreeuw2001,oliveira}. Structural nonseparability is naturally not limited to entanglement of just two degrees of freedom. Thus recently creation of a nonseparable, tripartite GHZ-like state of path, polarization, and transverse modes of a laser beam has been demonstrated in an experiment \cite{khoury-GHZ}.

An example of a successfully implemented quantum information protocol based on classical entanglement in optical fields with structured polarization is teleportation between different degrees of freedom.  Quantum information transfer between the spin and the optical angular momentum degree of freedom including transfer of two-photon quantum correlations has been experimentally demonstrated in Ref.~\cite{nagali} with vortex beams.  Using classical entanglement in radially polarized optical beams, quantum teleportation between path and spatial degrees of freedom has recently been reported and provides a novel method of distributing information between different transmission channels \cite{szameit}.  The required Bell measurement has been accomplished using an optical incarnation of the controlled-NOT gate (two Sagnac interferometers with a polarizing beamsplitter) followed by the Hadamard gate (another beamsplitter), and, finally, by a projective measurement in the  basis of four states spanning the joint Hilbert space of the two entangled cebits. Another interesting experimental demonstration of the quantum advantage using vortex beams is the implementation of a quantum game in the context of the prisoners dilemma \cite{khoury-games}.

Nonseparability between polarization and spatial degrees of freedom in cylindrical vector beams has been successfully used to accomplish various quantum and classical communication tasks. In optical communication it can be used, e.~g., for routing \cite{szameit} and for information encoding with higher density and less cross-talk \cite{milione}. In quantum communication, classical light nonseparable in polarization and spatial modes can be employed for the charaterization of a quantum channel \cite{forbes}, which is particularly relevant for free-space channels and replaces resource-intensive quantum state tomography. One can extend the quantum channel characterization to higher dimensions by using entanglement between wavelength and spatial degrees of freedom \cite{mabena}. In quantum key distribution, it has been demonstrated in a proof-of-principle experiment that the possibility to encode logical qubits into nonseparable states of polarization and spatial modes of the same photon allows to dispense with a traditional shared polarization reference frame and provides an additional security mechanism \cite{souzaQKD}. Another interesting aspect of structural nonseparability in vector fields has been discussed in Ref.~\cite{beating}. It is possible to engineer optical vector fields with the degree of nonseparability that oscillates as a function of propagation distance. The nonseparability dynamics occurs in free space under unitary conditions and can be realized both for coherent light and for single photons. This property of classical entanglement can find applications ranging from quantum key distribution to microscopy and laser material processing (for more comprehensive account see \cite{beating}).
The more detailed description of quantum information with classical optics and classical entanglement in optical communication is beyond the scope of the current review and the interested reader is referred to the cited work. 

From applications of classically entangled light in quantum technologies, we pick up high-precision measurements. In the subsequent section, we concentrated on use of the radially polarized classically entangled light in polarization metrology \cite{leuchs2014} and kinematic sensing \cite{leuchs2015}, which are based on correlations between spatial and polarization degrees of freedom and beautifully illustrate different concepts discussed above. Note that a much wide range of tasks related to polarization metrology can be accomplished exploiting classical entanglement. Degree of nonseparability is directly linked to the vector character of the optical field so that measuring of this nonseparability can be used to test for scalar or vector nature of the field and the corresponding entropy of entanglement is directly linked with the average degree of polarization \cite{forbes}. Not only the vector space of spatial degrees of freedom can be delpoyed to form classical entanglement for polarization applications. Recently, the polarization-frequency nonseparability has been used in measurements of a depolarization strength of materials \cite{fade}.

\subsection{High precision measurements using classical entanglement in radially polarized beams}

\begin{center}
\begin{figure}[h]
\begin{center}
\includegraphics[width=0.9\linewidth]{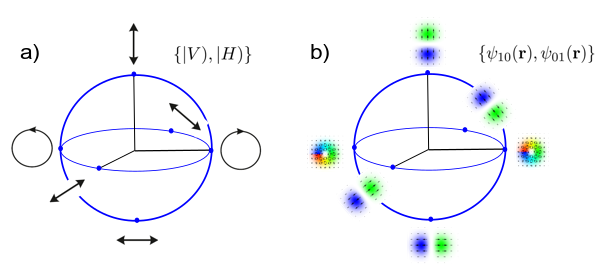} \end{center}
\caption{ (color online) Schematic visualization of a) the polarization Poincare sphere
representation of the binary Hilbert space of polarization cebit  $\{ \vert V ), \vert H )\}$ and b) Poincare sphere
representation of first-order spatial modes Hilbert space  $\{ \psi_{10}({\bf r}), \psi_{01}({\bf r})\}$ (see \cite{leuchs2014}).  }
\label{poincare}
\end{figure}
\end{center}

\begin{center}
\begin{figure}[h]
\begin{center}
\includegraphics[width=0.9\linewidth]{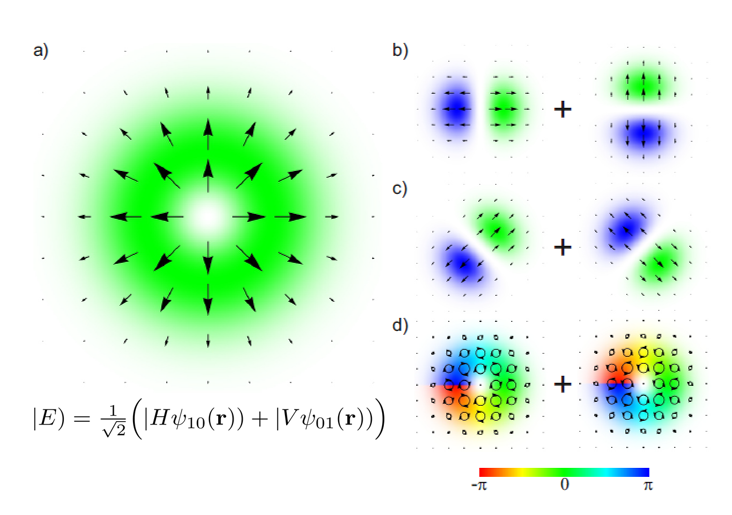} \end{center}
\caption{ (color online) Illustration of the polarization parallelism (modified from \cite{leuchs2014}). Radially polarized beam (a) with the electric field (\ref{Evector}) can be decomposed into different linear superpositions of the basis vectors in the product spaces $\{ \vert V ), \vert H )\}$ and $\{ \psi_{10}({\bf r}), \psi_{01}({\bf r}\})$: 
(b) $\vert E) =  \frac{1}{\sqrt{2}} \left [ \vert H \psi_{10}({\bf r})) + \vert V\psi_{01}({\bf r}) \right )]$; 
(c) $\vert E) =  \frac{1}{\sqrt{2}} \left [ \vert D_+ \psi_{+}({\bf r})) + \vert D_{-}\psi_{-}({\bf r})  \right )]$, 
where $D_{\pm} = \frac{1}{\sqrt{2}} \left ( \vert H ) \pm \vert V )\right )$ are diagonal polarizations and 
$\psi_{\pm} =\frac{1}{\sqrt{2}} \left ( \psi_{10} \pm \psi_{01}\right ) $ are diagonal HG spatial modes; 
(d) $\vert E) =  \frac{1}{\sqrt{2}} \left [ \vert C_L \psi_{R}({\bf r})) + \vert C_R \psi_{L}({\bf r})  \right )]$, 
where $C_{L,R} = \frac{1}{\sqrt{2}} \left ( \vert H ) \pm i \vert V )\right )$ are left and right circular polarizations and 
$\psi_{L,R} =\frac{1}{\sqrt{2}} \left ( \psi_{10} \pm i \psi_{01}\right ) $ are left and right circular HG spatial modes. The color scale (bottom) gives the phase of the electric field. }
\label{pol-parallel3}
\end{figure}
\end{center}

A lucid example of ``polarization parallelism'', analogous to quantum parallelism used in quantum computation, is the use of classical entanglement between spatial and polarization degrees of freedom in Mueller matrix polarimetry \cite{{simon-mueller},leuchs2014}.
Consider  the product Hilbert space of polarization and spatial degrees of freedom as introduced in sec.~\ref{hilbert}, only now the spatial Hilbert space is spanned not by the basis vectors of beam position, 
but by the basis vectors $\{\psi_{10}({\bf r}), \psi_{01}({\bf r})\}$,
where $\psi_{nm}({\bf r}), n,m=0,1$ is the Hermit-Gauss (HG) solution of the paraxial wave equation of the order $N=n+m=1$, i.~e., the first order HG spatial modes (see Fig.~\ref{poincare}). That is, the four-vector $\vert \Psi )$ of Eq.~(\ref{4vector}) will represent in such Hilbert space the electric field $\vert E )$ of a light beam nonuniformly polarized in transversal plane, specifically, radially polarized light beam. Defining the states of polarization- and spatial-cebits as $\vert 0,1 )$ for vertical, horisontal polarizations and $\psi_{10}({\bf r}), \psi_{01}({\bf r})$, respectively, we can write the electric field four-vector of the radially-polarized beam in a familiar form of a Bell state:
\begin{eqnarray}\label{Evector}
\vert E) =  \frac{1}{\sqrt{2}} \Big ( \vert H \psi_{10}({\bf r})) + \vert V\psi_{01}({\bf r})   ) \Big )=\frac{1}{\sqrt{2}} \Big ( \vert 00 ) + \vert 11) \Big )
\end{eqnarray}
in a combined Hilbert space of polarization $\{ \vert V ), \vert H )\}$ and spatial $\{\psi_{10}({\bf r}), \psi_{01}({\bf r})\}$ degrees of freedom (Fig.~\ref{pol-parallel3}). In a conventional Mueller matrix polarimetry, the detection scheme cannot resolve spatial and polarization degrees of freedom simultanuosly. Therefore, from four-vector description we need to revert to a more general representation using equivalent of the density matrix ${\bf \Gamma}=\vert E )( E \vert$ with $\vert E )$ given in Eq.~(\ref{Evector}). In the language of cebits $\vert 0,1 )$ it will again have the form of a Bell state density matrix:
\begin{eqnarray}\label{matrix} {\bf \Gamma}=\vert E )( E \vert = \frac{1}{2} \left (
\begin{array}{cccc}
	1&0&0&1 \\
	0&0&0&0 \\
	0&0&0&0 \\
	1&0&0&1 
\end{array}\right ).
\end{eqnarray}
Now if the detection scheme is not capable of resolving the spatial degrees of freedom, that would mean tracing out the unobserved degrees of freedom in Eq.~(\ref{matrix}), leaving us with a sensible $2 \times 2$ polarization coherence matrix $\rho_{\rm pol}$, 
the reduced matrix of ${\bf \Gamma}$. It can be represented in terms of Stokes parameters $S_j$ mentioned above as
\begin{eqnarray} \label{rho-pol}
\rho_{\rm pol} = \frac{1}{2} \sum_{j=0}^{3} S_j \sigma_j,
\end{eqnarray}
where $\sigma_j$ are conventional Pauli matrices. Correspondingly,
\begin{eqnarray} \label{stokes-rho-pol}
S_j = \tr \left [ \rho_{\rm pol} \sigma_j \right ].
\end{eqnarray}
For the radially polarized beam, $\rho_{\rm pol}= \rho_{\rm spa} = {\cal I}$, where $\rho_{\rm spa}$ is the reduced spatial density matrix and ${\cal I}$ is the $2 \times 2$ identity matrix. In classical polarization optics it means that the radially polarized beam is completely unpolarized, ${\bf S}= (S_0, S_1, S_2, S_3)^T=(1, 0, 0, 0)^T$. Although the light is radially polarized, the $\rho_{\rm pol}$ is a reduced matrix and this corresponds to measuring the global Stokes parameters of the beam as a whole, similarly as we discussed around Eq.~(\ref{class-Bell}) in sec.~\ref{hilbert}. This is also in accordance with the conclusion of the previous sec.~\ref{coh-pol}: polarization and entanglement are complimentary quantities, maximal entanglement corresponds to completely unpolarized light.

In a conventional Mueller matrix measurement
setting \cite{mueller}, an either transmissive or scattering material sample (the object) is illuminated
with a light beam (the probe) prepared in four different polarization states, $\alpha \in \{0,1,2,3 \}$, in a temporal
sequence. From the analysis of the polarization of the light transmitted or scattered, the optical properties of the object can be inferred.
Thus in a standard polarimetry, only one degree of freedom can be resolved in the measurement and 
using Eq.~(\ref{rho-pol}), (\ref{stokes-rho-pol}), one can obtain the following expression for the Stokes operators at the output of the measurement set-up \cite{leuchs2014}:
\begin{eqnarray}\label{mueller-conv}
S_j^{\rm out} (\alpha) = \sum_{k=0}^3 M_{jk} S_k (\alpha), \qquad j=0, 1, 2, 3,
\end{eqnarray} 
where $S_j^{\rm out}(\alpha)$ and $S_k (\alpha)$ are the Stokes parameters of the output and input beams in polarization state $\alpha$, respectively, and $M_{jk}$ denotes the unknown elements of the Mueller matrix $M$, which we are to determine. Then, conventionally, a linear system of 16 equations and 16 unkowns is constructed from Eq.~(\ref{mueller-conv}) and Muller matrix inferred.

In the novel setting using classical entanglement \cite{leuchs2014}, the object is
probed only once with one light beam of radial polarization, as opposed to four differently
polarized beams. Then, the light transmitted or reflected by the object is analyzed both in
polarization and in spatial degrees of freedom by means of suitable polarization and spatial mode selectors. Specifically, the polarization of the beam is used to actually probe the object and the
spatial degrees of freedom are used to post-select the polarization state of the light: this is the main idea
presented in \cite{leuchs2014}. This scheme outperforms conventional Mueller polarimetry because the
radially polarized beam carries all polarizations at once in a classically entangled state, thus
providing for a sort of ‘polarization parallelism’ (Fig.~\ref{pol-parallel3}). For this, density matrix and Stokes parameters from Eq.~(\ref{rho-pol}), (\ref{stokes-rho-pol}) are reformulated as two-degrees-of-freedom (TDoF) quantities:
\begin{eqnarray} \label{rho-pol-2}
\rho = \frac{1}{4} \sum_{j,k=0}^{3} S_{jk} \left ( \sigma_j \otimes \sigma_k \right )
\end{eqnarray}
where $\sigma_{j,k}$ are conventional Pauli matrices. The TDoF Stokes parameters are
\begin{eqnarray} \label{stokes-rho-pol-2}
S_{jk} = \tr \left [ \rho \left ( \sigma_j \otimes \sigma_k \right ) \right ], 
\end{eqnarray}
which are classical counterparts of the two-photon Stokes parameters
introduced in \cite{stokes2}. In \cite{stokes2}, the two polarization qubits are encoded in two
separated photons. The important difference, specific for classically entangled case, is that in (\ref{rho-pol-2}) the polarization cebit and the spatial cebit are encoded in the same radially polarized beam of light. Therefore, the two-degrees-of-freedom Stokes parameters give the intra-beam correlations between polarization and spatial degrees of freedom \cite{loudon}. In order to measure these
correlations, a special detection scheme has been derived in \cite{leuchs2014}, capable of resolving both degrees of freedom. We will concentrate here on the conceptual side and refer the reader interested in experimental details to \cite{leuchs2014}. 

For the radially polarized beam
represented by (\ref{matrix}),  the two-degrees-of-freedom Stokes parameters take the particularly simple form:
\begin{eqnarray} 
S_{jk} = \lambda_j \delta_{jk}, \qquad \left \{ \lambda_j \right \}^3_{j=0} = \left \{ 1,1,-1,1\right \}.
\end{eqnarray}
The output Stokes parameters then are given by:
\begin{eqnarray} 
S_{jk}^{\rm out} = \tr \left [ \rho^{\rm out} \left ( \sigma_j \otimes \sigma_k \right ) \right ] = M_{jk} \lambda_k .
\end{eqnarray}
As $\lambda_k =1$ for $k\in\left \{0,1,3\right \}$ or $\lambda_k =-1$ for $k=2$, the two-degrees-of-freedom Stokes parameters enable a direct measure of the Mueller matrix elements:
\begin{eqnarray}
M_{jk} = \left\{ 
\begin{array}{rr}
	- S^{\rm out}_{jk}, & {\rm for} \quad k=2, \\
S^{\rm out}_{jk}, & {\rm for} \quad k\neq 2.
\end{array} \right .
\end{eqnarray}
This means that the Mueller matrix of an object can be obtained from the measurement of the 16
two-degrees-of-freedom Stokes parameters $S^{\rm out}$, with a single radially polarized input beam, allowing the
performance of single-shot full polarimetry. This is the polarization parallelism enabled by classical entanglement.
For all practical applications where the optical properties of the sample change rapidly with time, the method of \cite{leuchs2014} presents an advantage over conventional Mueller matrix polarimetry \cite{mueller}. The drawback of the scheme is a more involved detection
setup compared to a conventional polarimetry and hence  potentially higher sensitivity to measurement errors.

Now consider the situation when an opaque object cuts across  the radially polarized beam. The motion of this object will modify the Stokes parameters of the beam and its spatial pattern. To account for the beam sensitivity to such moving obstacle, we can generalize Eq.~(\ref{Evector}) allowing for different weightings in the superposition \cite{leuchs2015}:
\begin{eqnarray}\label{Evector2}
\vert E) =  \lambda_1 \vert H \psi_{10}({\bf r})) + \lambda_2  \vert V\psi_{01}({\bf r}) ).
\end{eqnarray}
For $\lambda_1, \lambda_2 \neq 0 $ the radially polarized beam is unpolarized and classically entangled. That means, polarization and spatial degrees of freedom are strongly correlated. Due to this entanglement between spatial and polarization modes, the spatial and polarization patterns of the non-uniformly polarized beam vary with time according to the instant position of the moving object. More precisely, the Stokes parameters $S_{1,2,3}$ will be rescaled by $\lambda_1 - \lambda_2 $ and components of  $S_{1,2,3}$ (the electric field of orthogonal $x,y$ polarizations and  $\lambda_1, \lambda_2 $) will become the functions of the instantaneous coordinates of the obstacle. Consequently, if a moving object is obstructing the beam, we can retrieve information about its position entirely from polarization characteristics with no need for spatially resolving measurements. This is another example of polarimetry based on polarization parallelism: one can infer the instantaneous trajectory of the moving object from single-shot measurements of the Stokes parameters and these can be performed at GHz rate allowing for high-speed kinematic sensing \cite{leuchs2015}. 

\section{Conclusions}

Quantum coherence and quantum coherent superpositions at the level of individual subsystems are behind general quantum correlations beyond entanglement \cite{lectures-gen-cor} in noisy separable systems. Optical coherence, polarization degree and degree of classical entanglement are intimately connected in case of quantum intra-system correlations in nonseparable states of optical vector fields, like non-uniformly polarized light \cite{eberly2011,eberly2016}. Both intra-system nonseparability (entanglement of different degrees of freedom in a single system) and quantum inter-system correlations in mixed separable states are genuinely quantum features. Despite initial scepsis, they proved to be useful quantum resources, provide important theoretical framework and can be used for fundamental demonstrations of, e.~g., quantum properties of light and in diverse applications \cite{adesso-review,streltsov-review,spreeuw,saleh2012,aiello2015}. The inherent quantumness present in these apparently classical states manifests itself differently and should be employed in its own way, compared to bone fide non-local entanglement. This is yet another occasion for us to re-think quantum-classical boarder, in particular in optics \cite{eberly2016,spreeuw2001}. It is very inspiring research area, which brings long established concepts of coherence, polarization properties, inter- and intra-mode correlations,
and mathematics of vector spaces to the service of quantum technologies.

\section*{Acknowledgements}
The long-term support of our research on non-classical correlations in light beams from the International Max Planck Partnership (IMPP) with Scottish Universities, Alexander von Humboldt Foundation, the 7th EU Framework Programme under FET, the Scottish Universities Physics Alliance (SUPA), and the Engineering and Physical Sciences Research Council (EPSRC) is gratefully acknowledged.  We thank N. P. Karapetkova for the help in preparation of Fig.~\ref{ent-distr} (left) and Fig.~\ref{entangBS}.
\end{document}